\documentclass{elsart}

\usepackage{amsmath,graphicx,amsfonts}
\usepackage{bm}

\setkeys{Gin}{width=0.7\textwidth} 

\begin{document}

      \begin{frontmatter}
    \title{Anomalous Thermostat and Intraband Discrete Breathers}
\author[Dresden,Mainz,Saclay]{S. Aubry} and \author[Dresden,Mainz]{R. Schilling}
\address[Dresden]{Max-Planck-Institut f\"ur Physik komplexer Systeme, N\"othnitzer Str. 38,
01187 Dresden, Germany}
\address[Mainz]{Institut f\"ur Physik, Johannes Gutenberg-
      Universit\"at D-55099 Mainz, Germany}
\address[Saclay]{Laboratoire L\'eon Brillouin (CEA-CNRS), CEA Saclay,
91191 Gif-sur-Yvette Cedex, France}

\begin{abstract}
We investigate the dynamics of a macroscopic system which consists
of an anharmonic subsystem embedded in an arbitrary harmonic
lattice, including quenched disorder. The coupling between both
parts is bilinear. Elimination of the harmonic degrees of freedom
leads to a nonlinear Langevin equation with memory kernels
$\mathbf{\Gamma}(t)$ and noise term $\bm{\zeta}(t)$ for the
anharmonic coordinates $\mathbf {q} (t) = {(q_\alpha (t))}$. For
zero temperature, i.e~for $\bm{\zeta}(t)\equiv 0$, we prove that
the support of the Fourier transform of $\bm{\Gamma}(t)$ and of
the time averaged velocity-velocity correlations functions
$\bm{K}(t)$ of the anharmonic system can not overlap. As a
consequence, the asymptotic solutions can be constant, periodic,
quasiperiodic or almost periodic, and possibly weakly chaotic. For
a sinusoidal trajectory $\bm{q}(t)$ with frequency $\Omega$ we
find that the energy $E_T$ transferred to the harmonic system up
to time $T$ is proportional to $T^{\alpha}$. If $\Omega$ equals
one of the phonon frequencies $\omega_\nu$, it is $\alpha=2$. We
prove that there is a zero measure set $\mathcal{L}$ such that for
$\Omega$ in its full measure complement $\mathcal{R} \backslash \mathcal{L}$,
it is $\alpha=0$, i.e.~there is no energy dissipation. Under certain
conditions  $\mathcal{L}$ contains a subset $\mathcal{L}'$ such that for
$\Omega \in \mathcal{L}'$ the dissipation rate is nonzero and may
be subdissipative $(0 \leq \alpha < 1)$ or superdissipative $(1 <
\alpha \leq 2)$, compared to ordinary dissipation $(\alpha=1)$.
Consequently, the harmonic bath does act as an anomalous
thermostat, in variance with the common belief that elimination of
a macroscopically large number of degrees of freedom always
generates dissipation, forcing convergence to equilibrium.
Intraband discrete breathers are such solutions which do not
relax. We prove for arbitrary anharmonicity and small but finite
coupling that intraband discrete breathers with frequency $\Omega$
exist for all $\Omega$ in a Cantor set $\mathcal{C}(k)$ of finite
Lebesgue measure. This is achieved by estimating the contribution
of small denominators appearing for $\mathbf{G} (t;\Omega)$,
related to $\mathbf{\Gamma}(t)$. For $\Omega\in\mathcal{C}(k)$ the
small denominators do not lead to divergencies such that
$\mathbf{G} (t;\Omega)$ is a smooth and bounded function in $t$.
\end{abstract}
\end{frontmatter}

\section{Introduction}
The analytical treatment of the dynamics of a macroscopic system
of $N$ particles is not possible, in general. But there are
situations which allow to simplify the dynamical description. We
will restrict ourselves to classical systems. One of the prominent
examples is Brownian motion. A big particle with mass $M$
interacts with a solvent. If the solvent particles have a mass $m$
much smaller than $M$ a time scale separation exists. Separating
the fast motion of the solvent particles from that of the big one,
called Brownian particle, a Langevin equation results for the
motion of the Brownian particle. The influence of the solvent
particles occurs through a friction and a fluctuating force. These
two are related to each other by the so-called fluctuation-
dissipation theorem of second kind. For a general system with time
scale separation, the Mori-Zwanzig projection formalism \cite{1,2}
allows to derive a Langevin equation for the slow variables. If
there is a single slow variable $A$, then one gets \cite{1,2}

\begin{equation}
\frac{d}{dt}A(t) + i\Omega A(t) + \int\limits_0^tdt'\,M(t-t')A(t')
= f(t)\label{eq:1}
\end{equation}

where $\Omega$ is a microscopic frequency, $M(t)$ a memory kernel
and $f(t)$ the fluctuating force. Elimination of a macroscopic
number of fast degrees of freedom always leads to memory effects.
If $M(t)$ relaxes much faster then the average of $A(t)$, one can
perform a Markov approximation, i.e. $M (t)\approx 2\gamma\delta
(t-t')$. The analytical calculation of $M(t)$ is not feasible, in
general. However, there exists a class of models for which the
exact elimination of a macroscopic number of degrees of freedom is
possible, even if there is no time scale separation. These models
we call Kac-Zwanzig models \cite{3,4}. They consist of a subsystem
coupled to a macroscopic bath of \textit{harmonic} oscillators.
The subsystem can be microscopic, i.e. the number of its degrees
of freedom is of order one or it can also be macroscopic with,
e.g. 10$^{23}$ degrees of freedom. Since the equations of motion
are linear in the oscillator coordinates, the harmonic degrees of
freedom can be microscopic, i.e.~the number of its degrees of
freedom is of order one or it can also be macroscopic with, e.g.
10$^{23}$ degrees of freedom. As a result, one obtains for the
coordinates of the subsystem a stochastic differential equation of
type of eq.\eqref{eq:1}. The memory kernel $M$ (which is a matrix
in general) can be expressed by an integral over the frequencies
of the harmonic bath. Its spectral properties strongly influence
the time dependence of $M$ and therefore that of the coordinates
of the subsystem, too. In the Langevin case where
 the spectrum is absolutely continuous with infinite support, the
subsystem will relax to equilibrium for all of its initial
conditions. In this case, the harmonic oscillators act as a normal
thermostat. However, if, e.g.~that spectrum contains a dense
discrete component as it occurs for harmonic systems with quenched
disorder, the situation may be different. Then, there exist
initial conditions, which will not relax to equilibrium, but may
converge to periodic, quasiperiodic or possibly weakly chaotic
solutions. This behavior is closely related to the existence of
spontaneous energy localization.\\

Spontaneous energy localization in large nonlinear discrete
systems is now known as an ubiquitous phenomena related to the
existence of Discrete Breathers (DBs). When a sufficiently large
amount of energy is injected locally in a system which may sustain
DBs and which is initially at low temperature, it is found that
although some part of this energy spreads over the system, a
substantial part of this energy may remain localized as a DB
\cite{ST88} over very long time. The discovery of this kind of
localized modes was so surprising because they existed in
systems with discrete translational invariance in space.
DBs also called \textit{Intrinsic
Localized Modes}(ILM) are spatially localized time periodic
solutions of discrete nonlinear Hamiltonian systems
\cite{CFK04,Aub97,FW98,FlachGo}. Since many of them are linearly
stable and thus can trap energy over long times, they should play
an essential role in energy relaxation in complex systems.

The existence of DBs has been rigorously proven by several methods
and for a variety of models with optical or acoustic phonons, with
or without gaps, in one or any dimension and without or with
randomness \cite{MA94,LSM97,Aub98,AKK01,Jam03}. All these
existence proofs require that the DB frequency and its harmonics
do not belong to the phonon spectrum (obtained by linearization of
the dynamical equations in the vicinity of the ground state).
Otherwise, the DB would be expected to radiate energy in the
phonon band and to decay. In addition,  the proofs \cite{MA94,LSM97,Aub98,AKK01,Jam03} only concern the existence
of \textit{out-of-band}  DBs, i.e. their frequencies and those of the higher harmonics
are in the band gaps of the phonon spectrum. For certain models
the existence of DBs was even proven before by
Alban\`ese and Fr\"ohlich \cite{AF88,AF91} and by
Alban\`ese, Fr\"ohlich and Spencer \cite{AFS88} under the requirement
of randomness. In that case  the eigenstates of the linear
eigenvalue problem are already localized (in dimensions larger than two
the randomness has to be large enough)
and therefore it is more intuitive that localized time periodic solutions
could persist in the presence of nonlinearities, although to prove this is
highly nontrivial

When the phonon spectrum is discrete, for example in a
strongly disordered system with Anderson localization, energy radiation of
linear phonons by a localized time periodic solution cannot occur.
Then, the existence of \textit{Intraband}  Discrete Breathers (IDBs) with
their frequency or some of their harmonics \textit{inside} the phonon
spectrum, might be a priori possible. Indeed, numerical evidence
\cite{KA99,KA00} suggested that such solutions do exist. More
precisely, it was conjectured that for each localized Anderson
mode, there is an associated family of IDBs with variable
frequencies (and amplitude), which converges to this Anderson mode
at the small amplitude limit. It was also conjectured that the
IDBs with respect to their frequency dependence do not form a
continuous family of solutions as for ordinary (out-of-band) DBs,
but are defined only on a fat Cantor set of frequencies  (i.e.
with nonvanishing Lebesgue measure). A forbidden frequency gap is
associated with each of the other frequencies of the linear
Anderson modes. These frequencies are dense in the linear phonon
spectrum but the width of these gaps drops to zero very fast as a
function of the spatial distance of the corresponding Anderson
mode of the center to the IDB. Then, the complementary Cantor set
of frequencies where the IDB exists is non void with a strictly
positive Lebesgue measure. Moreover, the measure density of this
Cantor set  is expected to go to full measure when the amplitude
of the IDB goes to zero that is when the IDB tends to be a linear
mode.

The IDB should not be confused with so-called embedded solitons in
discrete lattice systems (see \cite{GFM04,YCM05} and references
therein). These are nonlinear excitations with a \textit{discrete}
frequency lying within the linear spectrum, whereas IDB take
frequencies from a set of full measure. Furthermore, in contrast
to IDB the existence of embedded solitons requires exponentially
localized linear modes.

The first proof of existence of such IDBs  for
a certain class of random models was done by Alban\`ese and Fr\"ohlich
 \cite{AF88,AF91} and by  Alban\`ese, Fr\"ohlich and Spencer \cite{AFS88}.
The models studied by these authors  are the \textit{random},
discrete  Schr\"odinger equation and the wave equation with a
cubic nonlinearity.  A perturbational approach was used to prove
that all periodic solutions of the linear equation can be
continued for small enough  $\lambda$, the strength of
nonlinearity. For arbitrary $\lambda$ it was proven that there
exist at least certain periodic solutions.  Transforming a
\textit{random}, nonlinear dynamical system without acoustic
phonons into Anderson space, i.e. to the space of the localized
eigenstates of the linear problem and restricting to nearest
neighbor interactions in Anderson space, Fr\"ohlich,  Spencer and
Wayne \cite{FSW} have even proven (under certain conditions) the
existence of localized \textit{almost periodic} solutions, i.e. of
KAM tori.

Besides the existence of IDBs, it was also conjectured in
\cite{KA99,KA00} the existence of families of multibreather states
which exhibit several energy peaks associated with different
Anderson modes. We may explain intuitively this conjecture by the
fact that the frequency of a nonlinear mode depends on its
amplitude. It is thus possible to tune the amplitude of an
arbitrary number of different Anderson modes in order they get the
same frequency. Despite nonlinear modes cannot be simply
superposed as in the linear case, a time periodic multibreather
solution can be constructed numerically. When, the number of IDBs
involved in the multibreather state is infinite, these
multibreather states are spatially extended and able to transport
energy by phase torsion \cite{CA97}. As a result, nonlinearity
restores the ability of the system to transport energy.

The aim of this paper is to provide an exact proof of these
conjectures for a \textit {special} class of  random models  where
the underlying physical arguments  appear explicitly. These
models, consisting of a microscopic or macroscopic anharmonic
subsystem coupled to a bath of harmonic oscillators, are designed
on one hand for obtaining simpler mathematical proofs and on the
other hand to contain nevertheless the essential features,
e.g.~for the trajectories $\bm{q}(t)$ of the anharmonic system,
which can be found in more general complex and realistic models.
Although these results do not provide a rigorous proof (in the
mathematical sense) for general and fully anharmonic systems, they
nevertheless make clearer the fundamental reasons for their
existence and thus support the earlier conjecture that they might
exist in general.

Particulary, we will derive for  rather general anharmonic
and harmonic subsystems(arbitrary dimension, with or
without disorder) an exact formula which relates the
energy dissipation rate ${\overline{{\dot{E}}_T}}$ of the
anharmonic system averaged over a time interval of length $T$ to
the Fourier transform ${\tilde{\bm\Gamma}}(\omega)$ of the
dissipation kernels $\bm{\Gamma}(t)$ and to ${\bm{\tilde{K}}}
(\omega)$, the Fourier transform of the velocity-velocity
correlation functions of the trajectories $\bm{q}(t)$. For finite
initial energy and bounded Hamiltonian from below it is
${\overline{\dot{E}}_T}\rightarrow 0$ for $T \rightarrow \infty$.
This will yield the nonoverlapping criterion for the support of
${\bm{\tilde{\Gamma}}}(\omega)$ and ${\bm{\tilde{K}}} (\omega)$
which will have consequences for the possible asymptotic behavior
of  $\bm{q}(t)$ for $t \rightarrow \infty$. Calculation of the
energy dissipation rate ${\overline{\dot{E}}_T}$ for a sinusoidal
force with frequency $\Omega$ will demonstrate the existence of
extreme sensitivity of the dissipated energy by that force on its
frequency $\Omega$. Compared to ordinary dissipation where $E_T
\sim T$, superdissipative and subdissipative behavior may occur
for which $E_T$ increases, respectively, faster and slower than
$T$. We will also show that there exists a set of full measure
such that $E_T$ is bounded for all $\Omega$ in that set, i.e.~no
energy dissipation exists at all.

The possibility of no energy dissipation allows the existence of
periodic solutions $\bm{q}(t+2 \pi/\Omega)=\bm{q}(t)$, also within
the phonon band. We will show that such IDB exist with frequencies
on a fat Cantor set. In that case relaxation to equilibrium is
prohibited such that the harmonic bath acts as an
\textit{anomalous} thermostat. This behavior is related to the
pioneering work of Fermi, Pasta and Ulam, which has given evidence
for nonergodic dynamics. New results confirm that ergodicity in
phase space of large and complex dynamical systems is not as
smooth as it was believed through Langevin bath theory. When
starting from initial states which are not at thermal equilibrium,
there are many \textit{islands} in phase space which may trap the
state of the system for  long time and could result in unusual and
slow relaxation processes toward the thermal equilibrium.

\section{Modeling the Intraband Discrete Breather Problem}

In order to keep our presentation selfcontained we review and
discuss properties of linear spectra which have been proven
earlier or are well-admitted in order to define a model and an
associated set of hypothesis which could match with the different
physical situations. Our purpose is to describe the main
characteristics which could be expected for large families of
nonlinear random models with linear spectrum involving both
discrete and continuous components. Continuous spectra occur only
in the thermodynamic limit, i.e.~for an infinite system.

In order to fix the ideas, let us consider as an example a random
model on a d-dimensional hypercubic lattice, for example a
Klein-Gordon type model with Hamiltonian
\begin{equation}
\mathcal{H}= \sum_i
\left(\frac{p_i^2}{2}+V_i(u_i)\right)+\sum_{\langle i,j \rangle }
 \frac{C}{2} (u_i-u_j)^2 \label{rnlham}
\end{equation}
where the momentum $p_i$ is the variable conjugate to the
displacement $u_i$ at site $i$. The local potential $V_i(u_i)$ is
random. For example it could have the form

\begin{equation}
V_i(u_i)=\frac{\omega_i^2}{2}u_i^2 +\frac{\kappa_3}{3} u_i^3
+\frac{\kappa_4}{4} u_i^4+... \label{locpot}
\end{equation}

where $\omega_i$ is chosen according to some probability law
$P(\omega)$. $\langle i,j \rangle $ denotes the bonds which
connect neighboring sites $(i,j)$ (counted once). Accordingly, the
sum in eq.\eqref{rnlham} is a double sum over all $i$ and $j$
which are nearest neighbors. $C$ is the constant of the harmonic
coupling between nearest neighbors, which is assumed to be
identical for all nearest neighbor bonds. The dynamical equations
are
\begin{equation}
\ddot{u}_i + V_i^{\prime}(u_i)+\sum_{j:i} C(u_j-u_i) =0
\label{dyneqg}
\end{equation}
$j:i$ denotes the sites $j$ which are nearest neighbors to $i$,
i.e.~the sum in eq.\eqref{dyneqg} is a single sum over $j$ which
are nearest neighbors of $i$. DBs are spatially localized time
periodic solutions of eq.\eqref{dyneqg}. Thus, their amplitude
must decay to zero at infinity. Consequently, we may approximate
this dynamical system by another dynamical one which obeys the
same eq.\eqref{dyneqg} for the few sites $i$ where the amplitudes
of the DB is not small and the linearized equations
\begin{equation}
\ddot{u}_i + V_i^{\prime \prime}(0) u_i+\sum_{j:i} C(u_j-u_i) =0
\label{dyneqgl0}
\end{equation}
for all the other sites $i$. Thus, the system becomes equivalent
to a finite  anharmonic system coupled linearly by local
interactions to a harmonic system (phonon bath). Note, we start
with a finite harmonic system and finally take the thermodynamic
limit. For a finite harmonic system the linear spectrum is always
discrete.

This kind of approximation strongly simplifies the investigation
of DBs. Assuming the existence of time periodic solutions (DBs) of
the {\it isolated} anharmonic system (i.e.~uncoupled with the
harmonic phonon bath) we will discuss under which conditions these
solutions could persist when the coupling to the phonon bath is
non vanishing, although not too large. This harmonic phonon bath
considered uncoupled to the anharmonic system, is characterized by
its frequency spectrum. When the frequency of the DB and its
harmonics do not belong to the frequency spectrum of the phonon
bath, the proof that the DB solutions can be continued to nonzero
coupling can be obtained as in \cite{MA94}. This will be performed
in the $5.$ section. When the frequency of the DB, one of its
harmonics or several belong to the frequency spectrum of the
phonon bath, it is essential to consider whether this phonon
spectrum is discrete or continuous.

\begin{figure}[h!]
\begin{center}
\includegraphics{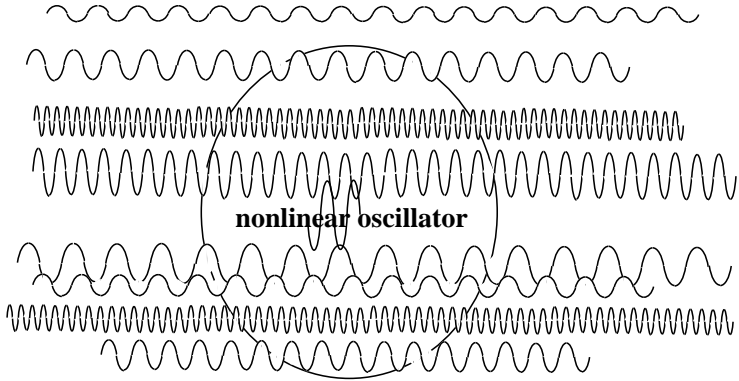}
\caption{The eigenmodes of the uncoupled phonon bath are extended
waves at the DB frequency. Then, at nonvanishing coupling periodic
oscillations of the anharmonic system should generate waves which
propagate and carry energy at infinity. As a result, the DB
solution cannot persist as a finite energy (spatially localized)
time periodic solution.} \label{fig1}
\end{center}
\end{figure}

\begin{figure}[h!]
\begin{center}
\includegraphics{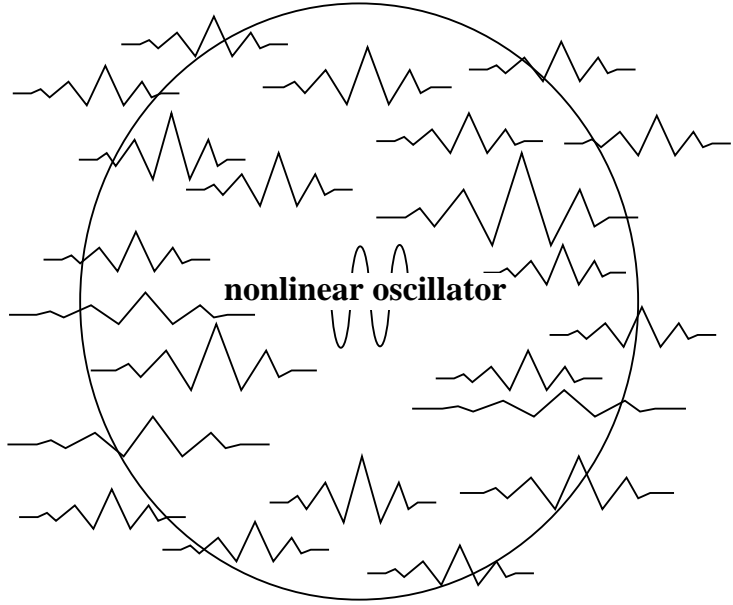}
\caption{The eigenmodes of the uncoupled phonon bath are spatially
localized at the DB frequency. No energy propagation is possible.
Then, the DB solution of the anharmonic system may persist at
finite coupling.} \label{fig2}
\end{center}
\end{figure}

Figs.~\ref{fig1} and \ref{fig2}, respectively, show possible
schemes representing situations where the linear phonons are
either spatially extended or localized at the DB frequency.

The model we propose is simply to consider the system as harmonic
in the regions where the amplitude of the DB is expected to be
small and to keep it anharmonic in the region where the solution
has a large amplitude. Of course, our model shall depend on the
location of the IDB or multiDBs we consider since it determines
the selection of the anharmonic subsystem.

This kind of simplification also allows to investigate the
relaxation of an arbitrary localized initial excitation. If the
finite initial energy is spread completely over the system the
displacements become arbitrary small and the linearization of
eq.\eqref{dyneqg} is allowed for large enough times. However, if
the spreading is incomplete, we are allowed to linearize
eq.\eqref{dyneqg} outside the localized region.

When $\{u_i\}$ is \textit {uniformly} small, the linearized
equations which yield the linear phonon spectrum are
\begin{equation}
\ddot{\epsilon}_i + \omega_i^2 \epsilon_i
 +\sum_{j:i} C(\epsilon_j-\epsilon_i) =0
\label{dyneqgl}
\end{equation}
which has time periodic solutions of the form
$\epsilon^{(\alpha)}_i(t) =\epsilon^{(\alpha)}_i
e^{i\omega_{\alpha} t}$. The eigenfrequencies $\omega_{\alpha}$
and eigenmodes
$\bm{\epsilon}^{(\alpha)}=\{\epsilon_i^{(\alpha)}\}$ obey the
eigenequation
\begin{equation}
\left(\mathbf{A_0\bm{\epsilon}^{(\alpha)}}\right)_i = \omega_i^2
\epsilon_i^{(\alpha)}+\sum_{j:i}C(\epsilon_j^{(\alpha)}-\epsilon_i^{(\alpha)})
= \omega_{\alpha}^2\epsilon_i^{(\alpha)} \label{eigmd}
\end{equation}
where $\mathbf{A}_0$ is the matrix of the second order variation
of the potential energy from its ground state. The
eigenfrequencies $\omega_\alpha$ should not be confused with the
frequencies $\omega_i$ of the small amplitude oscillations in the
local potential $V_i(u_i)$.

\subsection{Linear Spectrum Properties}

It has been proven \cite{AL} that the linear spectrum of
eq.\eqref{eigmd} is discrete and its eigenvectors (Anderson modes)
are square summable and exponentially localized with probability
one, when either the spatial dimension is less than three or the
coupling is not too large. At larger coupling and when the
dimension of the lattice is at least three, it is believed that
there exist mobility edges in the spectrum (although no rigorous
proof is available up to now). There are two mobility edges which
determine a central interval where the spectrum is continuous. The
corresponding eigenmodes are extended and non square summable. In
two intervals limited by the edge of the spectrum, the spectrum is
discrete and the corresponding eigenmodes are square summable and
exponentially localized.

More generally, we may consider nonlinear models on arbitrary
discrete lattices where the small amplitude solution leads to the
eigenvalue equation

\begin {equation}
\mathbf{A} \mathbf{x}_\alpha = E_\alpha \mathbf{x}_\alpha
\label{eigeq}
\end{equation}
for an arbitrary  positive  bounded  self-adjoint linear operator
$\mathbf{A}$ operating in the Hilbert space
$\mathbf{x}=\{x_{\mathbf{n}}\}$ where $\mathbf{n}$ represent the
sites of an infinite discrete lattice $\mathcal{L}$ in $1,2,3...$
dimensions (or any discrete network), with the  standard Hermitian
product $\langle\mathbf{x}|\mathbf{y}\rangle= \sum_{\mathbf{n}}
x_{\mathbf{n}}^* y_{\mathbf{n}}$ and $||\mathbf{x}||=
(\langle\mathbf{x}|\mathbf{x}\rangle)^{1/2}$.  Since we consider
physical systems which have to be linearly stable, we must assume
the spectrum to be positive or zero and then, the eigenvalues
$E_\alpha=\omega_{\alpha}^2$ are the square of the eigenmode
frequencies. We may also consider systems with acoustic phonons
where the spectrum is gapless and contains the translation mode
$x_{\mathbf{n}}\equiv a$ associated with zero eigenfrequency which
is always extended.

Standard mathematical theories decompose the spectrum
$\mathcal{S}$ of such a self-adjoint bounded linear operator
$\mathbf{A}$ in three parts: the absolutely continuous spectrum,
the discrete spectrum and the singular continuous spectrum
\cite{mathtxt}. The characteristics of the spectrum manifests
itself in finite size systems in the limit of large size $N$ as
explained now. The eigenvalue equation restricted to a finite
subset $\mathcal{N}$ of $N$ indices of $\mathcal{L}$ corresponds
to the diagonalization of a finite self-adjoint matrix
\begin{equation}
\mathbf{A}^{(\mathcal{N})}\mathbf{x}_{\alpha}^{(\mathcal{N})}
 = E_{\alpha}^{(\mathcal{N})} \mathbf{x}_{\alpha}^{(\mathcal{N})}
 \label{eigN}
\end{equation}
which yields $N$ orthogonal and normalized eigensolutions
$\mathbf{x}_{\alpha}^{(\mathcal{N})}$ with compact support
$\mathcal{N}$ corresponding to  real eigenvalues. Taking
$N\rightarrow +\infty$ such that the finite support tends to cover
the infinite lattice $\mathcal{N}\rightarrow \mathcal{L}$, the
nonvanishing accumulation points for the weak topology
$\mathbf{x}_\alpha$\footnote{A vector is said to converge to a
limit vector for the weak topology when each of its components
converge to a limit.} of the eigensolutions of finite systems
$\mathbf{x}_{\alpha}^{(\mathcal{N})}$ are eigensolutions of
eq.\eqref{eigeq} for an eigenvalue $E_\alpha$, the corresponding
limit of $E_{\alpha}^{(\mathcal{N})}$. The nonvanishing limit
solutions are necessarily square summable with a nonzero norm
smaller or equal to $\|\mathbf{x}_{\alpha}^{(\mathcal{N})}\|=1$
and thus can be normalized.

This set of square summable eigensolutions $\mathbf{x}_{\alpha}$
with eigenvalues $E_{\alpha}$  determines the discrete spectrum of
$\mathbf{A}$. Since  $\mathcal{L}$ is countable, the discrete set
of normalized eigenvectors is also countable (or finite). The
inverse rate of the exponential decay of the eigenmodes determines
the so called localization length which is a smooth function of
the eigenfrequency and diverges when approaching mobility edges.
In systems with acoustic phonons in 3 dimensions and more, it is
believed that the spectrum becomes continuous in a nonvanishing
positive interval close to zero and then there is only one
mobility edge.

Extended states (e.g. plane waves) are not square summable and
their norm diverges proportionally to the system size $N$. These
eigenstates can be nevertheless obtained as nonvanishing
accumulations points of
 $\sqrt{N}\mathbf{x}_{\alpha}^{(\mathcal{N})}$.
This set of nonsquare summable solutions and their corresponding
eigenvalues  of the eigenvalue equation \eqref{eigeq} are usually
called pseudo eigensolutions  and pseudo eigenvalues. They
corresponds to the absolutely continuous spectrum of $\mathbf{A}$.
They form uncountable continua.

The limits of the other finite size eigenstates which behave
spatially  differently for example which develop a fractal
structure at large scale or which exhibit algebraic decay at
infinity, determine the remaining part of the spectrum which is
the singular continuous spectrum of $\mathbf{A}$. Although there
are examples of operators with purely singular continuous spectrum
(for example with quasiperiodic potentials), their contribution to
the spectrum of most operators is expected to be marginal,
generally, and then it reduces to few isolated specific points of
the spectrum (corresponding to the mobility edges).

The spectrum of an operator $\mathbf{A}$ (for example random) is a
closed subset $\mathcal{S}$ of the real axis. It might be purely
absolutely continuous (for example for operators with spatial
periodicity) or purely discrete (for example for operators in 1d
with randomness or higher dimension with strong disorder) or it
might decompose into two disconnected subsets $\mathcal{S}_d$ and
$\mathcal{S}_{ac}$ where the spectrum is discrete and absolutely
continuous, respectively, separated by mobility edges. This
situation is supposed to occur for random operators in 3
dimensions and more at weak enough disorder\cite{KA00}.

\subsection{The anticontinuous Limit}
Considering now again, the nonlinear system \eqref{dyneqg} at the
anticontinuous limit $C=0$, it simply becomes a collection of
uncoupled nonlinear oscillators with Hamiltonian $H_i=\frac{p_i^2}
{2} +V_i(u_i)$ and thus is trivially integrable. It is then
convenient to represent each anharmonic oscillator with the
standard action-angle variables $I_i,\theta_i$. Then, the local
Hamiltonian $H_i(I_i)$ is only a function of its action $I_i$, and
its frequency is $H^{\prime}_i(I_i)=\omega_i(I_i)$. Note, that
$\omega_i(I_i)$ should not be confused with the frequencies
$\omega_i$ in eq.\eqref{locpot}. Time periodic solutions at
frequency $\Omega$ are obtained by choosing each oscillator $i$
either at rest or oscillating at frequency $\Omega=\omega_i(I_i)$
if there exists $I_i$ such that this equality may be fulfilled.
The solution corresponding to a Discrete Breather at site $i$ and
at frequency $\Omega$  is obtained by choosing all the oscillators
$j\neq i$ at rest, except oscillator $i$ which is oscillating at
frequency $\Omega$.

At $C=0$, the linear spectrum of eigenfrequencies defined by
eq.\eqref{eigmd} is the {\it closure}
$\mathcal{S}_L=\overline{\bigcup_{i} \{\omega_{i}\}}$ of the set
of linear frequencies. Considering a DB solution at the
anticontinuous limit at site $i$ and at a frequency $\Omega$ which
is nonresonant with the linear spectrum (that is $p\Omega \notin
\mathcal{S}_L$ for any integer $p$), and fulfills  the condition
that $d\omega_i(I_i)/dI_i =H^{\prime \prime}_i(I_i)\neq 0$ at
$\omega_i(I_i)=\Omega$, it was proven by the implicit function
theorem \cite{MA94}, that this  solution can be continued as a
spatially localized solution for $C\neq 0$ belonging to a
nonvanishing interval.

When $\Omega \in \mathcal{S}_L$, the proof for the DB existence
given in \cite{MA94} which  requires the invertibility of
$\mathbf{A}_0-\Omega^2\mathbf{1}$ breaks down. However, numerical
investigations suggested the existence of stable IDBs with
frequency $\Omega \in \mathcal{S}_L$ although they are not
continuous functions of the model parameters. Otherwise, this
numerical work also suggested the existence of intraband multisite
breathers called multibreathers. A direct proof of existence of
IDB in fully anharmonic models is likely possible (as an extension
of the present work) but it should be more complex for technical
reasons  than in the simplified model we
propose. We think, however, that this model captures the essential
physical features of general nonlinear random models.

\subsection{Defining a model for IDBs}

We denote by $\mathbf{q}=\{q_{\alpha}\}$ the set of variables
which describe the coupled anharmonic oscillators and by
$\mathbf{p}$ their conjugate variables. This subsystem is
described by an anharmonic Hamiltonian

\begin{equation}
\mathcal{H}_0(\mathbf{p},\mathbf{q})=
\frac{1}{2}\mathbf{p}^2+V(\mathbf{q}) \label{exham}
\end{equation}

Without restricting generality the masses are chosen to be equal
to one. The single anharmonic oscillator where $\mathbf{q}$ is a
scalar, is the simplest but a special case. This anharmonic part
is  linearly  coupled to an infinite harmonic system described by
the set of variables $\bm{u}=\{u_i\}$ and conjugate momenta
$\bm{p}=\{p_i\}$ where  the indices $i$ are nodes of an arbitrary
network. The harmonic Hamiltonian is $\mathcal{H}_{\rm
harm}(\{u_i,p_i\})= \frac{1}{2} \bm{p}^2+ \frac{1}{2} \bm{u}^t
\bm{M}\bm{u}$ where $\bm{M}$ is a  bounded self-adjoint operator
and $t$ denotes transposition. Again, the masses can be taken
equal to unity without lack of generality. The total Hamiltonian
takes the form
\begin{equation}
\mathcal{H} =\mathcal{H }_0(\mathbf{p},\mathbf{q})+
 \mathcal{H}_{\rm harm}(\{u_i,p_i\}) - \lambda  \mathbf{q}^t \mathbf{C}\bm{u}
\label{phb0}
\end{equation}
where the matrix of coupling constants is
$\mathbf{C}=\{C_{\alpha,i}\}$ and $\lambda$ is introducted for
convenience. These coupling constants are short ranged that is
$\sum_i |C_{\alpha,i}|^2 <+\infty$.

It is more convenient to describe the harmonic system as a
collection of noninteracting harmonic oscillators $\nu$ which
corresponds to its normal modes. The normal mode coordinates of
the harmonic oscillators of the phonon bath are now the
coefficients $u_{\nu}$ of the expansion of $\{u_i\}$ with respect
to the normalized eigenmodes $\bm{u}^{(\nu)}$ with frequency
$\omega_{\nu}$ which fulfills $\bm{M}\bm{u}^{(\nu)}
=\omega_{\nu}^2 \bm{u}^{(\nu)}$. They are now coupled to the
anharmonic system  by the set of coefficients
$\mathbf{C}_{\nu}=\bm{u}^{(\nu)t} \bm{C}$ or explicitly
$C_{\nu,\alpha} =\sum\limits_i u^{(\nu)}_i C_{\alpha, i}$.

As we assumed above, the set of coupling constants
$\mathbf{C_{i}}=(C_{\mathbf{\alpha},i})$ between $\mathbf{q}$ and
$u_i$ are  short ranged. In most physical situations, they decay
exponentially, that is  there exists two positive constants $K$
and $\gamma$ such that
\begin{equation}
 |\mathbf{C}_{\mathbf{i}}| < K e^{-\gamma |\mathbf{i}|} \quad .
\label{shrg}
\end{equation}
Thus, we may assume  the  stronger $\textit{L}_1$ condition
$\sum_{\mathbf{i}} |\mathbf{C}_i| < +\infty$ without loosing much
generality.

Then, we obtain a total Hamiltonian of the form
\begin{equation}
\mathcal{H} =\mathcal{H }_0(\mathbf{p},\mathbf{q}) - \lambda
\sum_{\nu} \mathbf{C}_{\nu}^t \mathbf{q}~ u_{\nu} + \sum_{\nu}
\left(\frac{1}{2}\dot{u}_{\nu}^2 +\frac{1}{2} \omega_{\nu}^2
u_{\nu}^2 \right) \quad .\label{phb}
\end{equation}

The reorganization energy $\mathcal{H}_R$  is the change in the
groundstate energy of the whole system under the constraint of a
static displacement $\bm{q}$ of the anharmonic oscillators. It is
obtained by minimizing the last two terms of the energy
(\ref{phb}) with respect to $u_{\nu}$ for fixed $\mathbf{q}$. This
readily yields
$\mathcal{H}_R(\mathbf{q})=-\frac{\lambda^2}{2}\mathbf{q}^t
\mathbf{K}_R\,\mathbf{q}$ where
\begin{equation}
K_{\alpha,\beta}^{(R)}=\sum_{\nu}
\frac{C_{\nu,\alpha}C_{\nu,\beta}} {\omega_{\nu}^2} < +\infty
\quad . \label{hypf1}
\end{equation}
In well-behaved physical models, this sum is always finite since
any \textit{finite} static displacement
 $\mathbf{q}$ should involve a \textit{finite}  reorganization energy
(and a finite force). This condition requires the assumption that
the positive definite series (\eqref{hypf1}) is always absolutely
convergent. For example, the case of models with acoustic modes
where one of the $\omega_{\nu}$ may vanish because there is always
a mode at zero frequency (which corresponds to the free motion of
the “center of mass” of the whole system). This feature may
cause a divergency for coefficients of series (\ref{hypf1}). But
in the limit of low frequency,  these (acoustic) phonon modes
decouple from the local degree of freedom describing the
anharmonic part of the system ($\lim_{\omega_{\nu}\rightarrow 0}
\mathbf{C}_{\nu}=0$) and thus  this reorganization energy
(\ref{hypf1}) remains finite since $\bm{C}_\nu$ is linear or even
of higher order in $\omega_\nu$ for $\omega_\nu \rightarrow 0$.
The reorganization energy $\mathcal{H}_R$ renormalizes the
potential energy of the anharmonic subsystem, i.e.~we can
introduce an effective energy
$\mathcal{H}_{eff}(\mathbf{p},\mathbf{q})=
\mathcal{H}_0(\mathbf{p},\mathbf{q})+\mathcal{H}_R(\mathbf{q})$.
Thus, the total Hamiltonian (\ref{phb}) may be written as
\begin{equation}
\mathcal{H} = \mathcal{H }_{eff}(\mathbf{p},\mathbf{q})+
\sum_{\nu} \left(\frac{1}{2}\dot{u}_{\nu}^2 +\frac{1}{2}
\omega_{\nu}^2 (u_{\nu}-\frac{\lambda \bm{C}^t_{\nu}
\bm{q}}{\omega_{\nu}^2})^2 \right) \quad ,\label{phb2}
\end{equation}

where $u_\nu(\mathbf{q}) = \lambda
\mathbf{C}_\nu^t\mathbf{q}/\omega_\nu^2$ is the harmonic
displacement minimizing the last two terms of \eqref{phb}, for
fixed $\mathbf{q}$. Since we did not specify properties of the
harmonic subsystem, which is considered as very general, it is now
necessary to introduce some assumptions at this level for ensuring
our model to be physically realistic. The density  of the positive
eigenfrequencies $\omega_{\nu}$ of the harmonic bath (called
density of states) is the limit of the positive measure defined as
$$d\mu(\omega)= \lim_{N\rightarrow +\infty} \frac{1}{N} \sum_{\nu} \delta(\omega-\omega_{\nu}) d\omega$$
In physical situations, it is generally an absolutely continuous
measure, dense on a finite interval (that is
$d\mu(\omega)=g(\omega) d\omega$ where $g(\omega)$ is a measurable
positive function). But this assumption is not necessary and
moreover does not characterize whether the phonon spectrum is
discrete or continuous.

For example, in spatially periodic harmonic systems the whole
phonon spectrum is absolutely continuous with extended eigenstates
which are planewaves. In random Anderson models in three
dimensions and  at weak enough disorder, the phonon spectrum
$\mathcal{S}$ may split into two disconnected sets
$\mathcal{S}=\overline{\mathcal{S}_{ac}\bigcup \mathcal{S}_d}$
which consists of a central interval $\mathcal{S}_{ac}$ where the
spectrum is supposed to be absolutely continuous (corresponding to
extended eigenstates) and the union $\mathcal{S}_d$ of two side
intervals where the spectrum is supposed to be discrete (and
corresponding to square summable eigenstates). These  intervals
are separated by mobility edges (isolated points). At large
disorder in 3d or at lower dimension for any disorder, the whole
spectrum is discrete and consists of a single interval
$\mathcal{S}_d$.

Actually, in all these examples, the density of states is an
absolutely continuous measure which does not determine a priori
whether or not the phonon spectrum is continuous or discrete. Note
also that in the case of random potentials, the density of states
depends only on the statistical properties of the potential and
not on the realizations. The characteristics of the spectrum
(absolutely continuous, singular continuous  or discrete) appears
in our model through the properties of the distribution of
coupling constants $\mathbf{C}_{\nu}$. They determine the
properties of the kernel which appears after elimination of the
harmonic variables and which shall be discussed later.

\section{Extended Langevin Theory}

Models for an anharmonic system coupled to a harmonic phonon bath
are ubiquitous in physics for describing relaxation processes, for
example chemical reactions within Kramers theory \cite{4,HTB90}.
However, in those applications the spectrum of the harmonic bath
was always assumed to be absolutely continuous. In that case,
standard calculations yield after elimination of the phonon bath
variables, extra damping terms for the anharmonic system (plus
Langevin random forces, if the temperature is not zero). When
there is a discrete spectrum, the results are totally different,
as we shall see now.

The well-known advantage of a coupling with a harmonic phonon bath
is that it is straightforward to eliminate the variables of the
phonon bath in order to get effective equations for the anharmonic
variables \cite{4} (we already took advantage of such a situation
for proving the existence of (interband) Discrete Breathers in
systems with acoustic phonons \cite{Aub98}).

\subsection{Derivation of a Langevin Equation}

Making use of eqs. \eqref{exham} and \eqref{phb}, the dynamical
equations are

\begin{eqnarray}
\ddot{\mathbf{q}}+V'(\mathbf{q}) - \lambda \sum_{\nu}
\mathbf{C}_{\nu}
u_{\nu} &=& 0 \label{anharmeq}\\
\ddot{u}_{\nu}  + \omega_{\nu}^2 u_{\nu} - \lambda
\mathbf{C}_{\nu}^t \mathbf{q} &=& 0 \label{harmeq}
\end{eqnarray}

where $V'(\mathbf{q})$ stands for the gradient of $V$ with respect
to $\mathbf{q}$. The second equation is linear and readily yields

\begin{equation}
u_{\nu}(t) = u_{\nu}^{(0)}(t) + \lambda
\frac{1}{\omega_{\nu}}\cdot \int\limits_{0}^{t}
\sin{\omega_{\nu}(t-\tau)}~
\mathbf{C}_{\nu}^t\mathbf{q}(\tau)\,d\tau \label{greene}
\end{equation}

where  $$u_{\nu}^{(0)}(t)=\left[\frac{ \lambda}{\omega_{\nu}^2}
\mathbf{C}_{\nu}^t\mathbf{q}(0) + \epsilon_{\nu}(0)\right] \cos
{\omega_{\nu}t }+ \frac{\dot{\epsilon}_{\nu}(0)}{\omega_{\nu}}
\sin  {\omega_{\nu}t }$$

is an arbitrary solution of  the homogeneous part of
eq.\eqref{harmeq} \footnote{Note that we choose the harmonic
displacement $ \epsilon_{\nu}(0) \cos {\omega_{\nu}t }+
\frac{\dot{\epsilon}_{\nu}(0)}{\omega_{\nu}} \sin  {\omega_{\nu}t
} $ with respect to the equilibrium position  $\frac{
\lambda}{\omega_{\nu}^2} \mathbf{C}^t_{\nu}\mathbf{q}(0)$ for
given initial coordinates $\mathbf{q}(0)$. }. If we assume the
phonon bath in thermal equilibrium, $\epsilon_{\nu}^{(0)}(t)$ is a
random variable due to thermal fluctuations and we have at
temperature $T$,
 $\langle \epsilon_{\nu}^{(0)}(0) \epsilon_{\nu}^{(0)}(t)\rangle
= \frac{k_B{T}}{\omega_{\nu}^2}\cos  \omega_{\nu}t$. Substituting
$u_{\nu}(t)$ from \eqref{greene} into \eqref{anharmeq}, we obtain

\begin{equation}
\ddot{\mathbf{q}}(t)+V^{\prime}(\mathbf{q}(t)) - \lambda^2
\int\limits_{0}^{t}\mathbf{G} (t-\tau) \mathbf{q}(\tau)\,d\tau -
\lambda^2 \sum_{\nu} \mathbf{C}_{\nu}(\mathbf{C}_{\nu}^t
\mathbf{q}(0)) \cos \omega_{\nu}t - \lambda\bm{\zeta}(t) =  0
\label{effenp}
\end{equation}
where the coefficients $G_{\alpha,\beta}(t)$ of matrix $\mathbf{G}
(t)$ are defined by
\begin{equation}
G_{\alpha,\beta}(t)= \sum_{\nu}
\frac{C_{\nu,\alpha}C_{\nu,\beta}}{\omega_{\nu}} \sin \omega_{\nu}
t \label{kernp}
\end{equation}
This series is absolutely convergent and yields a smooth function
of $t$ because of hypothesis \eqref{hypf1} and $
\omega_{\nu}<\omega_c< \infty$:
$$ \sum_{\nu} \frac{|C_{\nu,\alpha}C_{\nu,\beta}|}{\omega_{\nu}}
< \omega_{c} \sum_{\nu}
\frac{\mathbf{C}_{\nu}^2}{\omega_{\nu}^2}<+\infty$$
$\bm{\zeta}(t)$ is the Langevin random force
\begin{equation}
\bm{\zeta}(t) = \sum_ {\nu}
\epsilon_{\nu}^{(0)}(t)\mathbf{C}_{\nu} \label{lgvf}
\end{equation}

It is convenient to rewrite eq.\eqref{effenp} as
\begin{equation}
\ddot{\mathbf{q}}(t) + V_{eff}'(\mathbf{q}(t))+ \lambda^2
\int\limits_{0}^{t}\mathbf{\Gamma}(
t-\tau)\mathbf{\dot{q}}(\tau)\,d\tau - \lambda\bm{\zeta}(t) =  0
\label{effenp1}
\end{equation}
where
\begin{eqnarray}
V_{eff}(\bm{q})&=& V(\bm{q}) - \frac{1}{2} \lambda^2
\sum_{\alpha,\beta} \Gamma_{\alpha,\beta}(0)
q_{\alpha} q_{\beta} \label{veff} \\
\Gamma_{\alpha,\beta}(t)&=& \sum_{\nu}
\frac{C_{\nu,\alpha}C_{\nu,\beta}}{\omega_{\nu}^2}  \cos
\omega_{\nu} t \label{kernp2}
\end{eqnarray}
and  $d\mathbf{\Gamma}(t)/dt= -\mathbf{G}(t)$. We note
$\mathbf{\Gamma}(0)=\mathbf{K}_R$ which corresponds to the
coefficients of the reorganization energy \eqref{hypf1}.
Eq.(\ref{effenp1}) has the form of the Langevin equation
(\ref{eq:1}), however with the nonlinear term
$V_{eff}'(\bm{q}(t))$.

The damping kernel $\mathbf{\Gamma}(t)$ is related to the random
force $\bm{\zeta}(t)$ by the general Langevin relation
\begin{equation}
\langle\zeta_{\alpha}(0)\zeta_{\beta}(t)\rangle = k_B T\sum_ {\nu}
\frac{C_{\nu,\alpha}C_{\nu,\beta}} {\omega_{\nu}^2} \cos
\omega_{\nu} t = k_B T \Gamma_{\alpha,\beta} (t)\label{lgrel}
\end{equation}

where $\langle \quad  \rangle$ denotes the canonical average with
respect to $\{\varepsilon_\nu (0)\}$ and $\{\dot{\varepsilon}_\nu
(0)\}$.

Definition \eqref{kernp2} shows that the damping kernel
$\mathbf{\Gamma}(t) = \int \bm{\tilde{\Gamma}} (\omega) e^{i\omega
t} d\omega$ is the Fourier transform of
$\bm{\tilde{\Gamma}}(\omega)$ with matrix elements which are
measures
\begin{equation}
\tilde{\Gamma}_{\alpha,\beta}(\omega)d\omega = d
\sigma_{\alpha,\beta}(\omega) =\pi \sum_{\nu}
\frac{C_{\nu,\alpha}C_{\nu,\beta}}{\omega_{\nu}^2}
(\delta(\omega-\omega_{\nu}) +\delta(\omega+\omega_{\nu})) d\omega
\label{posmeas}
\end{equation}

Actually, measure  $\sigma_{\alpha,\beta}(\omega)$  may be written
in a more compact way as the result of an integration in the
complex plane
\begin{equation}
\sigma_{\alpha,\beta} (\omega)= \int_{0}^{\omega}
d\sigma_{\alpha,\beta}(\omega^{\prime})= \frac{1}{2}\frac{1}{2\pi
i}  \oint_{\mathcal{C}(\omega)}
 <\bm{C}_{\alpha}|\frac{1}{\bm{M}(z-\bm{M})} |\bm{C}_{\beta} > d z
\label{posmeas1}
\end{equation}
on a closed contour $\mathcal{C}(\omega)$ which encompasses the
part of the  spectrum of the positive linear phonon operator
$\bm{M}$ on the real axis which is smaller than $\omega^2$. Here
it is $\bm{C}_\alpha =(C_{\alpha,i})$. Note that in the case of
random potentials, this measure does depend on the realization of
the disorder unlike the density of states. This positive measure
$d\sigma_{\alpha,\beta}(\omega)$ can be split into three
components which are respectively absolutely continuous, discrete
and  singular continuous measures. eq.\eqref{posmeas1} proves that
this decomposition precisely corresponds to the decomposition of
the {\it spectrum} $\mathcal{S}$ of operator $\bm{M}$ which
describes the phonon bath, into its absolutely continuous
$\mathcal{S}_{ac}$, discrete $\mathcal{S}_d$ and singular
continuous parts $\mathcal{S}_{sc}$. Thus, the support of each of
these measure components on the real $\omega$ axis is independent
of $\alpha$ and $\beta$. Consequently, the properties of the
Fourier transform measure of the memory kernel
$\bm{\tilde{\Gamma}}(\omega)$  are directly related to the
spectral properties of the phonon bath.

\subsection{Relation between Phonon Spectrum and Kernel Properties}

We now show how the properties of the kernel $\bm{\Gamma} (t)$ and
of its Fourier transform $\bm{\tilde{\Gamma}}(\omega)$ are related
to the properties of the phonon bath. Actually, the Fourier
transform eq.(\ref{posmeas}) of the kernel $\bm{\Gamma}(t)$ is not
a priori a smooth function but is a positive measure
$\bm{\tilde{\Gamma}} (\omega)d\omega$
 which very generally can be decomposed into three parts:
the absolutely continuous part $\bm{\tilde{\Gamma}}_{ac}(\omega)$
where $\bm{\tilde{\Gamma}}_{ac}(\omega)$ is a positive measurable
function of $\omega$. The discrete part
$\bm{\tilde{\Gamma}}_{d}(\omega)$ which is a sum of Dirac
functions, and the singular continuous part
$\bm{\tilde{\Gamma}}_{sc}(\omega)$ is  simply the rest. These
different components are directly related to the corresponding
components of the  spectrum of the phonon bath. We will discuss
this relationship only qualitatively, without attempting any
rigor. The decomposition of the positive measure
$\bm{\tilde{\Gamma}}(\omega)d\omega$ into these three parts
implies a corresponding decomposition $\mathbf{\Gamma}
(t)=\mathbf{\Gamma}_{ac}(t)+ \mathbf{\Gamma}_d(t) +
\mathbf{\Gamma}_{sc}(t)$.Their qualitative time dependence will be
discussed below for the absolutely continuous and the discrete
part.

\begin{itemize}

\item[(i)] \textit{Absolutely Continuous Phonon spectrum}\\

\noindent For the eigenvalues $\omega_{\nu} \in \mathcal{S}_{ac}$
which belong to the absolutely continuous part of the phonon
spectrum, the eigenvectors $\bm{u}^{(\nu)}$ corresponding to
$\omega_{\nu}$ are extended, their components
$u_{\mathbf{i}}^{(\nu)}$ of the normalized state are proportional
to $1/\sqrt{N}$ where $N$ is the number of sites $i$ of the
harmonic part of our system  which is supposed to go to infinity.
Since we assumed that the coupling with the phonon bath
$\mathbf{C}_i$  is  local that is \eqref{shrg} is fulfilled, we
can write

\begin{equation}
\mathbf{C}_{\nu}\sim \frac{\mathbf{c}_{\nu}}{\sqrt{N}}
\label{abcspt}
\end{equation}

\noindent where $\mathbf{c}_{\nu}$ is a finite number independent
on $N$ for $N \rightarrow \infty$. Moreover, the extended
eigenvector $\bm{u}^{(\nu)}$ and consequently $\mathbf{c}_{\nu}$,
depend on a continuous parameter (for example the wave vector). An
important consequence is that the average of $\mathbf{c}_{\nu}$
for $\omega_{\nu}\in [\omega-\delta \omega,\omega+\delta \omega]$
in the limit of an infinite system and small $\delta \omega$ is a
well-defined measurable  function of $\omega$

\begin{equation}
\langle\mathbf{c}_{\nu}\rangle_{\omega_{\nu}=\omega} =
\mathbf{c}(\omega) \label{abcspt1}=(c_\alpha (\omega)) \quad .
\end{equation}

\noindent Thus, the contribution of  the absolutely continuous
phonon spectrum to the Fourier transform eq.(\ref{posmeas}) of the
Kernel $\bm{\tilde{\Gamma}} (\omega)d\omega$ is nothing but  its
absolutely continuous part $\bm{\tilde{\Gamma}}_{ac}(\omega)$. If
$\mu(\omega)$ is the density of states of the absolutely
continuous phonon spectrum the absolutely continuous measure is
given by

\begin{equation}
\tilde{\Gamma}_{\alpha,\beta}^{(ac)}(\omega)d\omega=\frac{c_{\alpha}(\omega)
c_{\beta}(\omega)}{\omega^2} \mu(\omega) d\omega \label{measureac}
\end{equation}\\

\noindent which implies

\begin{equation}
\Gamma_{\alpha,\beta}^{(ac)}(t)=\Gamma_{\alpha,\beta}^{(ac)}(-t)=\int_{\omega\in\mathcal{S}_{ac}}
\frac{c_{\alpha}(\omega) c_{\beta}(\omega)}{\omega^2} \cos (\omega
t) \mu(\omega) d\omega . \label{kernpc}
\end{equation}\\

\noindent Since $\mu(\omega)$ is absolutely integrable,
$\mathbf{\Gamma}_{ac}(t)$ decays as $1/t$ or faster, due to the
Riemann-Lebesgue lemma. This holds e.g. for a spatially periodic
system in any dimension and for a three- or higher-dimensional
system with disorder which is not too strong.\\

\item[(ii)] \textit{Discrete Phonon Spectrum}\\

\noindent For the eigenvalues  $\omega_{\nu} \in \mathcal{S}_d$
which belong to the discrete part of the phonon bath,
$\bm{u}^{(\nu)}$ is a localized square summable eigenmode. Then,
the scalar product $\mathbf{C}_{\nu} =\bm{u}^{(\nu)t} \bm{C}$ is
finite. Considering an arbitrary closed interval $\mathcal{I}_d
\subset \mathcal{S}_d$, we require a stronger convergence than
(\ref{hypf1}) for the discrete case:

\begin{equation}
\sum_{\omega_{\nu}\in \mathcal{I}_d }
 \frac{|\mathbf{C}_{\nu}|}{\omega_{\nu}}
=B(\mathcal{I}_d) < +\infty \quad .\label{hypf2}
\end{equation}\\

\noindent Actually, this assumption should be fulfilled in most
realistic
physical situations with a discrete spectrum.\\

\noindent Generally, the eigenmodes $\bm{u}^{(\nu)}$ decay
exponentially with respect to $|i|$ and there is a uniform bound
on the localization length provided $\omega_{\nu} \in
\mathcal{I}_d$ (provided this interval does not contain the
mobility edges). If we assume the coupling constants
$\mathbf{C}_{\mathbf{i}}$ fulfill \eqref{shrg}, then
$|\mathbf{C}_{\nu}|= |\bm{u}^{(\nu)t} \bm{C}|$ decays
exponentially as a function of $d_{\nu}$, the distance from the
``center of mass'' of mode $\nu$. Then, since $\omega_{\nu}$ is
bounded from below by a positive nonvanishing number, series
\eqref{hypf2} converges. Note, however, that even for a system
with acoustic phonons the series \eqref{hypf2} may exist because
the translational invariance implies at least
$\mathbf{C}_{\nu}\sim\omega_{\nu}$ for $\omega_{\nu}\rightarrow
0.$ However, in principle, there is no need for exponential
localization,
 the eigenmodes could decay  with some algebraic law providing
they are square summable.\\

\noindent The contribution of  the discrete spectrum of the phonon
bath to the Fourier transform measure (\ref{posmeas}) is nothing
but the discrete part $\bm{\tilde{\Gamma}}_{d}(\omega)$ which is a
sum of Dirac functions\\

\begin{equation}
\tilde{\Gamma}_{\alpha,\beta}^{(d)}(\omega)d\omega
 =\pi\sum_{\nu, \omega_\nu \in \mathcal{S}_d} \frac{C_{\nu,\alpha}C_{\nu,\beta}}{\omega_{\nu}^2} (\delta(\omega-\omega_{\nu})
+\delta(\omega+\omega_{\nu})) d\omega  \quad .\label{measured}
\end{equation}

\noindent This immediately leads to
\begin{equation}
\Gamma_{\alpha,\beta}^{(d)}(t)=
\Gamma_{\alpha,\beta}^{(d)}(-t)=\sum_{\nu,
\omega_{\nu}\in\mathcal{S}_d}
\frac{C_{\nu,\alpha}C_{\nu,\beta}}{\omega_{\nu}^2} \cos
\omega_{\nu} t \quad . \label{kernpd}
\end{equation}

\noindent Because of  \eqref{hypf1}, $\mathbf{\Gamma}_d(t)$ is an
absolutely convergent discrete sum of cosine functions.
Consequently, in general it is an almost periodic function which
does not  decay to zero at infinity but oscillates between a lower
and an upper bound. The situation of a pure discrete spectrum
occurs in disordered systems,
in $1d$ or $2d$ or in higher dimensions with strong enough disorder.\\

\item[(iii)] \textit{Singular Continuous Phonon Spectrum}\\

\noindent By definition, the singular continuous spectrum is the
rest of the phonon spectrum after removing the discrete and the
absolutely continuous part. If there is a singular continuous part
in the phonon spectrum, it simply appears as the singular
continuous component $\bm{\tilde{\Gamma}}_{sc}(\omega)$ of
$\bm{\tilde{\Gamma}}(\omega)$. In standard models which are
spatially periodic or  random, there is no singular continuous
spectrum but there are special models \cite{KST83,OPRS83} which are
neither spatially periodic nor random (for example quasiperiodic)
and where the whole spectrum is singular continuous. In such
models, for $\omega_{\nu} \in \mathcal{S}_{sc}$, the eigenvectors
$\bm{u}^{(\nu)}$ are not square summable though are not uniformy
extended as for example planewaves. They have a behavior at
infinity which is not universal.   For example, they might slowly
decay   with power laws. or they might not decay but with
intermittent and sparse peaks  with possibly fractal distribution
and zero density in average. We shall not much
discuss this exceptional situation in this paper.\\
\end{itemize}

If the disorder in $3d$ systems is weak enough a mixed situation
occurs with an absolutely continuous part in a central interval
and a discrete one in the  two intervals above and below the
mobility edges. Then, $\bm{\tilde{\Gamma}}(\omega)$ is the sum of
a discrete part and an absolutely continuous component with
supports which are the corresponding frequency intervals. When
there are acoustic phonons in a random model, the spectrum is
still fully discrete in one and likely two dimensions but with a
localization length which diverges at zero frequency. In three
dimensions, the low frequency phonon modes are extended. At larger
frequency beyond a mobility edge, the spectrum is discrete. In all
these  cases, the density of states remains a smooth measure
indistinctly the spectrum is absolutely continuous or discrete.

There are other situations we shall not consider here where the
spectrum may be absolutely continuous but where the support is a
fat Cantor set \cite{Hof76}(then $\bm{\tilde{\Gamma}}_{(ac)}(\omega)$ is not a
continuous function of $\omega$) or when the spectrum is discrete
but is not dense on finite intervals (for example in smooth
quasiperiodic potentials) or when the spectrum is purely singular
continuous (quasiperiodic square potentials \cite{KST83,OPRS83}). We shall
not discuss here these exceptional situations.

The Markov approximation is the simplest approximation which
consists in assuming that the spectrum of the phonon bath is only
absolutely continuous with infinite support and the coupling
constants $c_{\nu}$ are uniform. Thus
$\bm{\tilde{\Gamma}}(\omega)= \bm{\tilde{\Gamma}}_{ac}(\omega)=
const$ and $\Gamma_{\alpha,\beta}(t)=
\Gamma_{\alpha,\beta}^{(ac)}(t)=2\gamma_{\alpha,\beta} \delta(t)$.
Then, eq. \eqref{effenp1}  becomes
\begin{equation}
 \ddot{\mathbf{q}}(t)+V_{eff}^{\prime}(\mathbf{q}(t))+ \lambda^2
\bm{\gamma}\mathbf{\dot{q}}(t) - \lambda \bm{\zeta}(t) =  0
\label{effenpL}
\end{equation}

which  contains a damping term proportional to the velocities and
a Langevin force. The solution of this equation at zero
temperature, ($\bm{\zeta} (t)\equiv0$), always converge to a
static solution corresponding to the system at equilibrium.  This
result is essentially the consequence of the absolutely continuous
phonon spectrum which extends from frequency zero to infinity.

\section{Energy Dissipation and Asymptotic Solutions}
Starting from an initial condition where the phonon bath is at
rest, i.e.~ $\varepsilon_\nu (0)\equiv 0$ and
$\dot{\varepsilon}_\nu (0) \equiv 0$ the energy provided by the
anharmonic subsystem to this phonon bath may be viewed as energy
dissipation for this subsystem. In the general case, the rate of
energy dissipation is found to be related  to the Fourier
transform of the velocity-velocity correlation function of the
trajectory instead of being proportional to its square velocity as
in the original Langevin case. At zero temperature when the fluctuating
random force in eq.\eqref{effenp} vanishes, the possible
asymptotic solutions are those which could persist without any
energy dissipation. Since the dissipation rate depends on a time
average,
 the role of the transient dynamics toward the asymptotic solution disappears.

Actually, the asymptotic solution might not be always a static
solution when
 the structure of the phonon spectrum is not uniform. We already know
  situations  when the phonon spectrum is absolutely continuous
  but bounded with an upper bound (and
possibly with frequency gaps where the density of states vanishes,
 where (interband) DBs are asymptotic solutions.
Then their fundamental frequency (and harmonics) necessarily
belong to gaps of the
 phonon spectrum \cite{ST88,Aub97}.

We also know that when the phonon spectrum is purely discrete that
there are time-periodic solutions (called intraband DBs)  with
fundamental frequencies  which
 belong to a certain fat Cantor set  (with finite measure) \cite{KA00}  dense
 in the  phonon spectrum but which does not contain any phonon eigenfrequencies.
 However, we do not know if they
could be asymptotic solutions for other initial conditions which
are not those of that solutions. We would like to comment in that
section on the relationship between vanishing energy dissipation
and the possible asymptotic solutions for arbitrary initial
conditions. This relationship depends sensitively on the
properties of the phonon spectrum.

Let us first investigate the energy transferred after a long time
to the phonon bath
 of a solution
$\bm{q}(t)$ of  eq.(\ref{effenp}) (or of the equivalent equation
(\ref{effenp1})) at zero temperature, i.e. for $\bm{\zeta}(t) =0$.
The consequence for the asymptotic solutions will be discussed in
the final subsection.\\

\subsection{Derivation of an Energy Dissipation Functional}

For time $T>0$ given, let us first define functions\\

\begin{equation}
g_{\alpha;T}(t)=\chi_T(t) \dot{q}_{\alpha}(t) \label{veldef}
\end{equation}

where $\chi_T(t)=1/\sqrt{T}$ for $0<t<T$  and $\chi_T(t)=0$ for
$t<0$ and $t>T$, i.e. $\chi_T(t)$ is the normalized characteristic
function on $[0,T]$. Its Fourier transform is

\begin{equation}
\tilde{g}_{\alpha;T}(\omega)
=\int_{-\infty}^{+\infty}g_{\alpha;T}(t)  e^{i \omega t} dt =
  \frac{1}{\sqrt{T}} \int_0^T \dot{q}_{\alpha}(t) e^{i \omega t} dt \label{velfor}
\end{equation}

Let us now consider the effective energy of the anharmonic part of
the system defined as

\begin{equation}
\label{ anhenerg} E(t) = \frac{1}{2} (\bm{\dot{q}}(t))^2
+V_{eff}(\bm{q}(t)) .
\end{equation}

Using eq.(\ref{effenp1}) with $\bm{\zeta}(t)=0$, (since zero
temperature is assumed) the time derivative of this energy becomes

\begin{equation}
\label{avdis} \dot{E}(t) = - \lambda^2 \int_0^t \bm{\dot{q}}(t)
\cdot \bm{\Gamma}(t-\tau) \bm{\dot{q}}(\tau) d\tau \quad .
\end{equation}

The average of this time derivative from time $0$ to $T$ is
\begin{eqnarray}
\frac{E(T)-E(0)}{T}&=& {\overline{\dot{E}}}_T  = - \lambda^2
\frac{1}{T}
\int_0^T dt \int_0^t \bm{\dot{q}}(t) \cdot \bm{\Gamma}(t-\tau) \bm{\dot{q}}(\tau) d\tau  \nonumber\\
&=& - \frac{1}{2}\lambda^2 \frac{1}{T}
\int_0^T dt \int_0^T \bm{\dot{q}}(t)\cdot \bm{\Gamma}(t-\tau) \bm{\dot{q}}(\tau) d\tau \nonumber\\
&=& - \frac{1}{2}\lambda^2 \int_{-\infty}^{+\infty} dt
\int_{-\infty}^{+\infty}
 \bm{g}_T (t)\cdot \bm{\Gamma}(t-\tau) \bm{g}_T(\tau) d\tau
 \label{avdis1}
\end{eqnarray}
where we used  $\Gamma_{\alpha,\beta}(t-\tau)=
\Gamma_{\beta,\alpha}(\tau-t)$ and definition (\ref{veldef}).
Then, using the inverse Fourier transform $g_{\alpha;T}(t) =
1/2\pi \int_{-\infty}^{+\infty} e^{-i\omega t}
\tilde{g}_{\alpha;T}(\omega) d\omega$ and
$\int_{-\infty}^{+\infty} e^{i(\omega+\omega^{\prime})t} dt =2\pi
\delta(\omega+\omega^{\prime})$,we obtain the energy dissipation
functional
\begin{eqnarray}
{\overline{\dot{E}}}_T&=& -\frac{\lambda^2}{4 \pi}
\int_{-\infty}^{+\infty} \bm{\tilde{g}}_T^{\star}(\omega) \cdot
 \bm{\tilde{\Gamma}}(\omega) \bm{\tilde{g}}_T(\omega)  d\omega  \nonumber \\
 &=&-\frac{\lambda^2}{4 \pi} \sum_{\alpha,\beta}  \int_{-\infty}^{+\infty}  \tilde{g}_{\alpha;T}^{\star}(\omega)
\tilde{\Gamma}_{\alpha,\beta}(\omega) \tilde{g}_{\beta;T}(\omega)
d\omega \quad .\label{avdis3}
\end{eqnarray}

Definition (\ref{kernp2}) of $\bm{\Gamma}(t)$ implies that its
Fourier transform $ \bm{\tilde{\Gamma}}(\omega) $ is a positive
semidefinite matrix for any $\omega$, Consequently we have $
\overline{\dot{E}}_T (t)\leq 0$. This result is physically obvious
because the energy variation $E(T)-E(0)= T \overline{\dot{E}}_T$
of the anharmonic part of the system has to be negative at any
time $T$ since the total energy must be conserved and there was
initially no energy in the phonon bath.

$E(T)$ is obviously bounded from below (for physical stability of
the model) and from above by the total finite energy which is
initially injected to the system. Consequently, we have $ \lim_{T
\rightarrow +\infty} \frac{E(T)-E(0)}{T}=0$.  Thus a necessary
condition for the asymptotic solutions is that

\begin{equation}
\lim_{T \rightarrow +\infty} \overline{\dot{E}}_T=0, \quad .
\label{limedot}
\end{equation}

The limit of the products   in (\ref{avdis3}) for $T\rightarrow
+\infty$ might not be a smooth function of $\omega$ anymore. For
example if $\bm{q}(t)$ is time periodic, it should exhibit Dirac
peaks. Considering the limit within  the topology of weak
convergence of measures, this limit can be defined as a positive
measure matrix
 $\tilde{K}_{\beta,\alpha}(\omega) d\omega$. Then, it is proven
 in  appendix \ref{AppA}, that as a measure, this limit is the Fourier transform of
 the velocity-velocity correlation functions $K_{\alpha,\beta}(\tau)=
{\overline{\dot{q}_\alpha (t) \dot{q}_\beta (t+\tau)}}= \lim_{T
\rightarrow +\infty} \frac{1}{T}\int_0^T
\dot{q}_{\alpha}(t)\dot{q}_{\alpha}(t+\tau) dt:$

\begin{equation}
\tilde{K}_{\alpha, \beta}(\omega) d\omega = \lim_{T\rightarrow
+\infty} \tilde{g}_{\alpha;T}^{\star}(\omega)
\tilde{g}_{\beta;T}(\omega) d\omega = \left(
\int_{-\infty}^{+\infty}  K_{\alpha,\beta}(\tau) e^{i \omega \tau}
d\tau\right)d\omega \label{corrf}
\end{equation}

\subsection{Energy Dissipation for a Sinusoidal Periodic Driving
Force} \label{edspdf}


For a given solution $\bm{q}(t)$ of eq. \eqref{effenp1} with
$\mathcal{\zeta} (t)\equiv 0$ we have derived (in subsection 4.1)
an exact formula for the averaged energy dissipation rate. We will
apply this formula to a sinusoidal trajectory $\bm{q}(t)= \bm{a}
\cos (\Omega t- \theta)$ with frequency $\Omega$ and arbitrary
phase $\theta$. We have ${\dot{\bm{q}}}(t)= - \bm{a} \Omega  \sin(
\Omega t-\theta)$ which can also be interpreted as a driving
force. The normalized characteristic function fulfills $\int
\chi_T^2(t) dt=1.$ Its Fourier transform

\begin{equation}
\tilde{\chi}_T(\omega)= \frac{1}{\sqrt{T}} \int_0^T e^{i\omega t}
dt = \frac{2}{\omega \sqrt{T}} \sin \frac{\omega T}{2} e^{i\omega
T/2} \label{charf}
\end{equation}

fulfills $\int |\tilde{\chi}_T(\omega)|^2 d\omega=2 \pi$,
$\lim_{T\rightarrow +\infty} |\tilde{\chi}_T(\omega)|^2 = 2\pi
\delta(\omega)$ (for $\omega=0$, we set
$\tilde{\chi}_T(0)=\sqrt{T}$) and $\lim\limits_{T \rightarrow
\infty} \tilde{\chi}_T (\omega) \tilde{\chi}^*_T (\omega')=0$ for
$\omega \neq \omega$. Then, we can write eq.(\ref{velfor}) as
$\tilde{g}_{\alpha;T}(\omega) = \frac{1}{2\pi} \int
\tilde{\dot{q}}_{\alpha}(\omega-\omega^{\prime})
\tilde{\chi}_T(\omega^{\prime}) d\omega^{\prime}$ where
$\bm{\tilde{\dot{q}}}(\omega) = \int \bm{\dot{q}}(t) e^{i\omega t}
dt = i  \pi \bm{a} \Omega \left( e^{-i\theta}
\delta(\omega+\Omega)- e^{i\theta} \delta(\omega-\Omega)\right)$.
This yields:

\begin{equation}
 \tilde{g}_{\alpha;T}(\omega)= i \frac{a_{\alpha} }{2} \Omega \left(e^{-i\theta} \tilde{\chi}_T(\omega+\Omega)- e^{i\theta} \tilde{\chi}_T(\omega-\Omega) \right)
 \label{perdrive}
 \end{equation}

Then the average frequency dependent damping over time $T$ for
that time periodic function $\bm{q}(t)$ follows from
\eqref{avdis3}

\begin{eqnarray} {\overline{\dot{E}}}_T&=& -\frac{\lambda^2 \Omega^2}{16 \pi} \int
a_{\alpha} \tilde{\Gamma}_{\alpha,\beta}(\omega) a_{\beta}
\Big[\tilde{\chi}_T(\omega+\Omega)|^2 + |\tilde{\chi}_T(\omega-\Omega)|^2 -\nonumber \\
 &-& \Big(e^{-2i\theta} (\tilde{\chi}_T (\omega+\Omega) \tilde{\chi}_T^{\star}(\omega-\Omega)+c.c.\Big)\Big]
 d\omega\, .
\label{freqd}
\end{eqnarray}

The average dissipation rate ${\overline{\dot{E}}}_T$ depends
sensitively on the damping kernel ${\tilde{\bm{\Gamma}}}(\omega)$
which itself depends on the phonon spectrum. We will discuss this
dependence of ${\overline{\dot{E}}}_T$ for $T \rightarrow \infty$
for an absolutely continuous and a discrete phonon spectrum.

\begin{itemize}
\item[(i)] {\it Absolutely Continuous Phonon Spectrum.}\\

\noindent As discussed in (i) of subsection 3.2,
${\bm{\tilde{\Gamma}}}(\omega)d\omega \equiv
{\tilde{\bm{\Gamma}}}_{ac}(\omega) d\omega$ is an absolutely
continuous measure where $\bm{\tilde{\Gamma}}_{ac}(\omega)$ is a
positive continuous function on its support which is  the
absolutely continuous spectrum $\mathcal{S}_{ac}$. It has been
assumed to be the union of one or several intervals. $
\bm{\tilde{\Gamma}}_{ac}(\omega)d\omega$ is vanishing  elsewhere.
Then, using the properties of $\tilde{\chi}_T (\omega)$ from
above, we get a result independent of the phase $\theta$

\begin{eqnarray}
{\overline{\dot{E}}}_{\infty}&=& -\frac{\lambda^2
\Omega^2}{8}\sum_{\alpha,\beta} \int a_{\alpha}
\tilde{\Gamma}_{\alpha,\beta}(\omega) a_{\beta}
\left(\delta(\omega+\Omega) + \delta (\omega-\Omega)\right) d\omega \nonumber \\
 &=& -\Big(\frac{\lambda \Omega}{2}\Big)^2 \bm{a}^t {\bm{\tilde{\Gamma}}} (\Omega) \bm{a}
  \label{freqd2}
\end{eqnarray}

where ${\tilde{\bm{\Gamma}}}(-\omega)={\tilde{\bm{\Gamma}}}
(\omega)$ has been used.\\

\noindent Consequently, there is energy dissipation, only, if the
driving frequency $\Omega$ or ($-\Omega)$ belongs to the
absolutely continuous spectrum $\mathcal{S}_{ac}$. In that case
the rate of the energy dissipation is well-defined and the energy
of the
phonon bath linearly increases in average  with respect to time.\\

\noindent If $\Omega$ belongs to the edge of one of the intervals
the Dirac delta functions in eq.\eqref{freqd2} will contribute
only half to the integral, i.e.~  ${\overline{\dot{E}}}_{\infty}$ is still a constant.
Accordingly, $E_T \sim T$ for $T \rightarrow \infty$. However, this is not true  anymore
if in one-dimensional systems there is a van Hove singularity. In that case  ${\tilde{\bm{\Gamma}}}
(\omega)$ diverges and a superdissipative behavior should be expected(see next subsection).


\noindent When  $\Omega \notin \mathcal{S}_{ac}$ we have
$\min_{\omega \in \mathcal{S}_{ac}} \ |\Omega-\omega| =
d_{ac}(\Omega) >0$ and consequently $|\tilde{chi}_T(\omega \pm \Omega)| <
\frac{2}{\sqrt{T}} \frac{|sin\frac{(\omega \pm
\Omega)T|}{2}}{d_{ac} (\Omega)}$. Then it comes out from
eq.\eqref{freqd} that $|{\overline{\dot{E}}}_T| < K/T$ where $K$
is some finite constant. Since
${\overline{\dot{E}}}_T=(E(T)-E(0))/T$, the effective energy
$E(T)$ is monotonously decaying towards $\lim_{T\rightarrow
+\infty} E(T) = E_{\infty}$, but remains bounded
from below at infinite time.\\

\item[(ii)] {\it Discrete Phonon Spectrum}\\

\noindent This case has been discussed in (ii) of subsection 3.2.
$ \bm{\tilde{\Gamma}}_d(\omega)d\omega$ is a discrete measure
(\ref{measured}). Substituting this into eq.\eqref{avdis3} and
taking eq.\eqref{perdrive} into account yields\\

\begin{equation}
{\overline{\dot{E}}}_T=- \frac{\lambda^2 \Omega^2}{8} \sum_{\nu}
a_{\nu} |e^{-i\theta} \tilde{\chi}_T(\omega_{\nu}+\Omega)-
e^{i\theta} \tilde{\chi}_T(\omega_{\nu}-\Omega)|^2 \label{freqd3}
\end{equation}

\noindent where $a_{\nu}=\bm{a}^t \bm{C}_\nu \bm{C}^t_\nu
\bm{a}/\omega_\nu^2=\sum\limits_{\alpha, \beta} a_\alpha C_{\nu,
\alpha} C_{\nu, \beta} a_\beta /\omega_\nu^2$ are {\it positive}
coefficients.

\noindent Since the positive series $\sum_{\nu} a_{\nu} < +\infty$
is absolutely convergent, we can label the countable set of
eigenfrequencies by positive integers $\nu \in \mathcal{N}$ such
that the sequence $a_{\nu}$ is monotone decaying ($a_{\nu}\geq
a_{\nu+1}$ for any $\nu$ ). Thus, the resonances $\nu$ are simply
ordered according to their strength. Then, $a_{\nu}$ can be
supposed to converge exponentially to zero as in standard physical
models with discrete spectrum. The $\Omega$-dependent energy
dissipation will be more subtle, compared to case (i) of an
absolutely continuous phonon spectrum.\\

\noindent When $\Omega \notin \mathcal{S}_d$ does not belong to
the discrete spectrum (which is the closure of the set of
eigenvalues), the same proof as above for $\Omega \notin
\mathcal{S}_{ac}$ yields again that there is no energy dissipation
and that the phonon bath energy remains bounded from below. When
$\Omega \in \mathcal{S}_d$ and  when there is a resonance that is
there exists $\nu_0$ such that we have precisely $\Omega=\pm
\omega_{\nu_0}$, then series (\ref{freqd3}) contains a term
$\tilde{\chi}_T(\omega \pm \Omega_{\nu_0})= \sqrt{T}$ which diverges when
$T\rightarrow +\infty$.  Consequently, $ {\overline{\dot{E}}}_T $
diverges  as $-T$ when $T\rightarrow +\infty$ while the energy
absorbed by the phonon bath $E_T$ diverges proportionally to
$T^2$. This result should be expected since we know that the
energy of an harmonic oscillator driven at
its resonant frequency diverge as a function of time as $t^2$.\\

\noindent However when $\Omega \in \mathcal{S}_d$ but $\Omega\neq
\pm \omega_{\nu}~ \forall \nu$, the situation is more subtle
because series (\ref{freqd3}) though it does not contain any
diverging term, contains terms with  denominators $|(\omega_{\nu}
\pm \Omega)|$ which  may be arbitrarily small and may cause
divergency. Inspired by the construction of the Liouville numbers
from rational numbers, we construct a subset of $\mathcal{S}_d$
which has full measure  and where this series is absolutely
convergent
for all $\Omega$ in that subset.\\

With ${\tilde{\chi}}_T(\omega)$ from eq.\eqref{charf} we get from
eq.\eqref{freqd3}

 \begin{eqnarray}
F(T)=E(T)-E(0)&=& T {\overline{\dot{E}}}_T=-
\frac{\lambda^2\Omega^2}{2}
\sum_{\nu} a_{\nu} \times \nonumber \\
\Big{|} \frac{\sin
(\omega_{\nu}+\Omega)T/2}{\omega_{\nu}+\Omega}e^{i((\omega_{\nu}+\Omega)T/2-\theta)}
&-& \frac{\sin
(\omega_{\nu}-\Omega)T/2}{\omega_{\nu}-\Omega}e^{i((\omega_{\nu}-\Omega)T/2+\theta)}~\Big{|}^2
\quad . \label{freqd4}
\end{eqnarray}\\

\noindent $F(T)$ is an infinite series of periodic functions of
$T$. If this series is absolutely convergent, its sum is an almost
periodic function of $T$ (in the sense of H.Bohr\cite{Bre01}) which is smooth
and bounded. A sufficient condition for absolute convergence  of
series (\ref{freqd4}) is that the positive series $ \mathcal{A}
(\Omega) =\sum_{\nu} a_{\nu}
\left(\frac{1}{(\omega_{\nu}+\Omega)^2}+\frac{1
}{(\omega_{\nu}-\Omega)^2}\right)$ is convergent.\\

\noindent In appendix B we will prove that there exists a set
$\mathcal{L}$ of zero Lebesgue measure such that the series
$\mathcal{A}$ is convergent for all $\Omega \in \mathcal{R}
\backslash \mathcal{L}$. This implies that there exists a set
$\mathcal{R} \backslash \mathcal{L}$ of full measure such that the
energy $-F(T)$ transferred to the phonon bath remains finite and
bounded for all times, provided $\Omega$ belongs to
$\mathcal{R}\backslash \mathcal{L}$. On the other side, it is also
proven in appendix B that there exists a subset $\mathcal{L}'$ of
$\mathcal{L}$ of zero Lebesgues measure such that $F(T)$ diverges
for $T \rightarrow \infty$, if $\Omega \in \mathcal{L}'$. In
addition it is proven that under some conditions for
$\{\omega_\nu\}$ an arbitrary eigenfrequency $\omega_\nu$ belongs
to $\mathcal{L}'$ with probability zero, only. This excludes $F(T)
\sim T^2 $ for all $\Omega \in \mathcal{L}'$. For $\Omega \in
\mathcal{R}\backslash \mathcal{L}$ it is $F(T) \sim T^0$ and $F(T)
\sim T$ for an {\it ordinary} dissipative Langevin bath with an
absolutely continuous spectrum. Therefore we may conjecture that
$F(T) \sim T^\alpha$ if $\Omega \in \mathcal{L}'$, with an
exponent $\alpha$ which may depend on $\Omega$. Compared to
ordinary dissipation where $\alpha =1$, situations where $1<\alpha
\leq 2$ may be called {\it superdissipative} and those where $0<
\alpha < 1$ {\it subdissipative}.

It has been shown that the classical dissipation kernel
$\bm{\Gamma}(t)$ determines the nonlocal action occurring in the
path integral representation of the quantum mechanical propagator,
after elimination of the harmonic d.o.f. \cite{Leggett}. The
nonlocal action describes quantum dissipation. Depending on the
low frequency behavior of the spectral density $J(\omega)$
(corresponding to ${\bm{\tilde{\Gamma}}}(\omega))$, i.e.~
$J(\omega) \sim \omega^s$, for $\omega \rightarrow 0$, qualitative
different quantum behavior is found. For instance, for a particle
in a double well coupled to a harmonic bath there can be ohmic
damping $(s=1)$, superohmic $(s>1)$ and subohmic $(0<s<1)$ damping
\cite{Legrev}. Since the $\Omega$-dependent dissipation discussed
above is not related to the low frequency behavior of
${\bm{\tilde{\Gamma}}}(\omega)$, the implication of the
subdissipative and superdisspative behavior
for quantum dissipation is not obvious.\\


In conclusion, we have obtained here a rigorous proof that when
the phonon spectrum is discrete and with the assumption of fast
enough decay of coefficients $a_{\nu}$ (which is generally
fulfilled in physical models), there is no energy dissipation  by
the phonon bath when it is driven (locally) by a time periodic
force for  \textit{most} choices of its  frequency $\Omega$ (i.e.
probability 1),  when $\Omega$ belongs to the phonon spectrum
$\mathcal{S}_d$ (the closure of the set of eigenvalues). This
result first sharply contrasts the properties of a Langevin bath
with absolutely continuous spectrum which is always dissipative
and second, gives evidence for the existence of IDB. We also
proved the existence of a zero measure subset $\mathcal{L}'$ of
frequencies for the driving force where the energy absorbed by the
phonon bath is unbounded as a function of time. In that case, the
phonon bath may be superdissipative or subdissipative.\\

\item[(iii)] {\it Mixed Case}\\

When the phonon spectrum contains both an absolutely continuous
component and a discrete component, energy dissipation is simply
given by the sum of these two contributions. Thus for having no
energy dissipation,  $\Omega$ has not to belong to
$\mathcal{S}_{ac}$ but may belong to the subset $\mathcal{R}
\backslash \mathcal{L}$ in $\mathcal{S}_d$ defined above  where
there is no energy dissipation. Of course, there is no energy
dissipation when $\Omega$ does not belong to any components of the
spectrum $\mathcal{S}_{ac}\bigcup \mathcal{S}_d$.
\end{itemize}

\subsection{Possible Asymptotic Solutions}

If one chooses a general initial condition $\bm{q}(0)$,
${\dot{\bm{q}}}(0),$ $\{\varepsilon_\nu (0) \}$, $\{
\dot{\varepsilon}(0)\}$ for the full system one may ask for the
asymptotic behavior of the solution $\bm{x}(t)=(\bm{q}(t)$,
$\{\varepsilon_\nu (t)\})$ of eq.\eqref{effenp1} for $t
\rightarrow \infty$. The asymptotic solutions of a trajectory
$\bm{X}(t) = (\bm{q}(t),\{\varepsilon_\nu(t)\})$ generated from an
arbitrary initial condition are defined as the trajectories in the
intersection $\bigcap_T \mathcal{T}_T$ for all time $T$ of the
closure of the trajectory $\mathcal{T}_T=
\overline{\bm{X}(t)_{t\geq T}}$ starting from  time $T$. This set
could be a fixed point corresponding to a static solution. It
could be an invariant circle. Then the asymptotic trajectory would
be time periodic (Discrete Breathers). It could be also an
invariant torus and then the asymptotic trajectory would be
quasiperiodic. We should also not exclude a priori situations were
the limit set is a strange attractor with chaotic trajectories. We
wish to discriminate about the possible limits which are allowed
in principle according to the properties of the phonon bath.

In the following we will discuss qualitatively which asymptotic
solutions may be possible. From the average energy dissipation
rate (eq.\eqref{avdis3}) we get a nonoverlapping criterion  for
the support of both positive measures
${\tilde{\bm{\Gamma}}}(\omega)d\omega$ and ${\tilde{\bm{K}}}(\omega)
d\omega$. This criterion, together with the generation of higher
harmonics, can be used to specify features of the asymptotic
solutions.

\begin{itemize}
\item[(i)] {\it Nonoverlapping criteria}\\

\noindent Equation (\ref{avdis3}) relates the rate of energy
dissipation after a finite time $T$ to the smooth function
$\tilde{g}_T(\omega)$. For $T \rightarrow \infty$, $\tilde{g}_T
(\omega)$ is related to the Fourier transform of the
velocity-velocity correlation function (\ref{corrf}) which does
not depend on the possible transient regime near the initial time
but  essentially on the asymptotic solution. For $T \rightarrow
\infty$, the rate of energy dissipation becomes\\

\begin{equation}
{\overline{\dot{E}}}_{\infty}=\lim_{T\rightarrow +\infty}
\overline{\dot{E}}_{T}= -\frac{\lambda^2}{ 4 \pi}
\sum_{\alpha,\beta} \int_{-\infty}^{+\infty}
\tilde{\Gamma}_{\alpha,\beta}(\omega)
\tilde{K}_{\beta,\alpha}(\omega)  d\omega \label{avdis5}
\end{equation}\\

\noindent The positive matrix of limit measures
$\bm{\tilde{K}}(\omega) d\omega =d \bm{\rho}(\omega)$ (Fourier
transform of the matrix of velocity-velocity correlation
functions) can be decomposed as usual in three components
(absolutely continuous, discrete and singular continuous)
$$d \bm{\rho}(\omega)=d \bm{\rho}_{ac}(\omega)+d \bm{\rho}_{d}(\omega)+d
\bm{\rho}_{sc}(\omega)$$

Then, strictly speaking, the integral in eq.(\ref{avdis5}) of
 the product of two positive generalized functions may be
 mathematically undefined when both  measures $\bm{\tilde{\Gamma}}(\omega)
 d\omega$ and $\bm{\tilde{K}}(\omega)  d\omega$ are not purely absolutely continuous.
Such a situation may occur for example when both of these measures
exhibits a Dirac component  at the same frequency $\omega_{\nu}$.
However, in that case, the physical situation is clear because the
anharmonic system would drive the phonon bath with a resonant time
periodic force component at frequency $\omega_{\nu}$. Then, the
energy which is provided to  the resonant harmonic oscillator
should grow as a function of time as $t^2$, faster than $t$. This
resonant situation would correspond to a superdissipative behavior
where $|{\overline{\dot{E}}}_{\infty}|=+\infty$ since the  energy
dissipation  per unit time would diverges as $t$ for $t\rightarrow
+\infty$. More generally, we have to ascribe an infinite positive
value to the integral (\ref{avdis5}) if its integrand is a product
of two singular continuous or discrete functions and their
supports overlap.\\

Since the support of a positive measure plays a crucial role in
the following and not all readers may be familiar with it we give
a brief definition. A support $\mathcal{K}_{ac}$ for the
absolutely continuous part $d {\bm\rho}_{ac}(\omega)$  can be
chosen as  the set of points where the derivative
$\frac{d{\bm\rho}(\omega)}{d\omega}$  is well-defined (for most
$\omega$) and is nonvanishing (it is positive). It is a measurable
set with nonzero measure. A support $\mathcal{K}_{d}$ (the
smallest) of the discrete component $\delta \bm{\rho}_d (\omega)=
\sum_n k_n \delta (\omega-\Omega_n)d \omega$ is the countable set
of points $\Omega_n$ where $k_n>0$. The support $\mathcal{K}_{sc}$
of the singular continuous measure $d\bm{\rho}_{sc}(\omega)$ can
be chosen as the set of points where
$\frac{d\bm{\rho}_{sc}(\omega)}{d\omega}$ diverges to infinity
(This limit is defined for most $\omega$ and is zero by
definition). It is a measurable set which is uncountable.\\

\noindent We assume that the support of $\tilde{\Gamma}_{\alpha,
\beta} (\omega) d \omega$ is the same for all $\alpha, \beta$.
Eq.\eqref{posmeas} shows that this holds if for given $\nu$ the
coupling constants $C_{\nu, \alpha}$ are nonzero for all $\alpha$.
This may be not true in presence of special spatial symmetries.
The same assumption is made the for the support of
$\tilde{K}_{\alpha, \beta} (\omega) d \omega.$ In generic models
all variables $q_\alpha (t)$, $\alpha= 1,2,3, \cdots$ are coupled
one with each other. Consequently, we assume that the support of
the Fourier transform of the correlation functions
${\overline{q_\alpha (t) q_ \beta (t + \tau)}}$ does not depend on
$\alpha, \beta$. This implies that $\{\tilde{K}_{\alpha \beta}
(\omega) d (\omega)\}$
have a common support.\\

\noindent The main result is that since the initial energy of the
anharmonic system is finite and the total energy is bounded from
below (for stability reasons), a necessary condition for the
asymptotic solutions, is that the average rate of energy
dissipation at time infinity given by the integral (\ref{avdis5})
be strictly zero. This implies the support of the positive
measures $\bm{\tilde{\Gamma}}(\omega) d\omega$ and
$\bm{\tilde{K}}(\omega)  d\omega$ do not overlap.\\

\item[(ii)] {\it Conjectures for the limit solutions} \\

Since ${\bm{\tilde{g}}}_T(\omega)$ for $T \rightarrow \infty$ is
{\it not} the Fourier transform of the velocity
${\bm{\dot{q}}}(t)$ which itself is directly related to the
Fourier transform ${\bm{\tilde{q}}}(\omega)$ of the trajectory
$\bm{q}(t)$, the support of ${\bm{\tilde{K}}}(\omega) d \omega$
may not yield information on ${\bm{\tilde{q}}}(\omega) d \omega$.
If $\bm{q}(t)$ is periodic, quasiperiodic or almost period
\footnote{Almost periodic functions are convergent series of
sine and cosine functions with a countable set of frequencies. Quasiperiodic functions
are a special case of almost periodic functions where
the involved frequencies are linear combinations of a finite number
of frequencies with integer coefficients} then $\bm{K} (t)$ is periodic,
quasiperiodic or almost periodic. In this case the support of
${\bm{\tilde{K}}}(\omega) d \omega$ is identical with that of
${\bm{\tilde{q}}}(\omega) d \omega$. However, if $\bm{q}(t)$ is
chaotic it is not clear how ${\bm{\tilde{q}}}(\omega) d \omega$ is
related to ${\bm{\tilde{K}}}(\omega) d \omega.$\\

\noindent If $\Omega$ is in the support of
${\bm{\tilde{q}}}(\omega) d \omega$, the nonlinearities in
eq.\eqref{effenp1} will generate all higher harmonics with
frequencies $\Omega p$, with $p$ an integer. Similarly, if
$\Omega_1$ and $\Omega_2$ are in the support, the nonlinearities
will generate higher harmonics with frequencies $n_1 \Omega_1 +
n_2 \Omega_2$, with $n_1$ and $n_2$ integers. We assume that this
also holds for the support of ${\bm{\tilde{K}}}(\omega) d \omega.$
Therefore we conjecture that because of nonlinearities of the
anharmonic system, the supports $\mathcal{K}_{ac}$ (resp.
$\mathcal{K}_{d}$,$\,$ $\mathcal{K}_{sc}$) have the following
invariance properties. If $\Omega_1 \in \mathcal{K}_{ac}$ and
$\Omega_2 \in \mathcal{K}_{ac}$  (resp. $\mathcal{K}_{d}$,$\,$
$\mathcal{K}_{sc}$),
 then we have $n_1\Omega_1 +n_2\Omega_2 \in \mathcal{K}_{ac}$
  (resp. $\mathcal{K}_{d}$,$\,$ $\mathcal{K}_{sc}$) for any pair of integers $n_1$ and
  $n_2$. A measurable set which has  this property is either the full real
axis or has a zero measure (proof in appendix \ref{AppD}).
Consequently, $\mathcal{K}_{ac}$ should be the whole real axis and
therefore should overlap the support of the phonon spectrum
$\mathcal{S}$. This generates energy dissipation. Consequently,
there should not exist any absolutely continuous component in the
Fourier spectrum of the velocity-velocity correlation function but
only  a discrete component or a singular continuous one (or
both).\\

\noindent In case of a discrete component there are two
possibilities. First, ${\bm{\tilde{K}}} (\omega) d \omega \sim
\delta (\omega - \Omega) d \omega$, i.e.  $\bm{K}(t)$ is periodic
with period $T=2 \pi/\Omega$ which is consistent with an
asymptotically periodic solution $\bm{q} (t+ T) = \bm{q}(t)$ for
$t \rightarrow \infty$. Second, ${\bm{\tilde{K}}} (\omega) d
\omega =\sum\limits_n {\bm k}_n \delta(\omega - \Omega_n) d
\omega$ with $\sum\limits_n m_n \Omega_n \neq 0$ for all integers
$\{m_n\}$, i.e. $\bm{K}(t)$ is quasiperiodic or almost periodic.
This again is consistent with an asymptotically quasiperiodic or
almost periodic solution $\bm{q}(t)$.\\

\noindent In order that an asymptotically periodic solution exists
its frequency $\Omega$ has not been in the support of
${\bm{\tilde{\Gamma}}} (\omega) d \omega$. This requires $\Omega
\neq \omega_\nu$ for all $\nu$. We have already shown (cf.~(ii) of
subsection 4.2) that this condition is not sufficient. This point
will be discussed in great detail in section 5. For an
asymptotically quasiperiodic or almost periodic solution a
necessary condition is $\Omega _n \neq \omega_\nu$ for and $n$ and
all $\nu$. We do not discuss this case but mention that the
existence of such solutions were proven for a special class of
models with quenched disorder \cite{FSW}. We expect that such
solutions also exist for the present model.\\

Since solutions which are neither periodic, nor quasiperiodic (or
almost periodic) are chaotic, the case of a singular continuous
component of ${\bm{\tilde{K}}} (\omega) d \omega$ can only be
related to asymptotically chaotic solutions. The fact that the
component is singular and not absolutely continuous (which has
already been excluded) may imply that the chaos is not fully
developed. This subtle point is also not discussed further.\\

\noindent Finally, the asymptotic solution can also be a constant
one, i.e.~$\lim\limits_{t \rightarrow \infty} \bm{q} (t)=
\bm{q}_*$, where $\bm{q}_*$ is a local minimum of $V_{\rm eff}
(\bm{q})$. In that case it is ${\bm{\tilde{q}}}_\infty (\omega) d
\omega = 2 \pi \bm{q}_* \delta (\omega)$ and ${\bm{\tilde{K}}}
(\omega) d \omega \sim \omega^2 \delta (\omega) d \omega$, s.t.
${\overline{\bar{\dot{E}}}}_\infty =\sum\limits_{\alpha, \beta}
\int \tilde{\Gamma} _{\alpha, \beta} (\omega) \tilde{K}_{\beta,
\alpha} (\omega) d \omega=0$. Here we used that
$\tilde{\Gamma}_{\alpha \beta} (0)$ exists. \\

This discussion shows that asymptotic solutions can be constant,
periodic, quasiperiodic (or almost periodic) and may be weakly
chaotic. As far as we know the existence of asymptotically weak
chaotic solutions of eq.\eqref{effenp1} has never been
investigated. We should also stress again that we have discussed
{\it possible} asymptotic solutions, only. We have no information
which of these asymptotic solutions will occur for given initial
conditions, e.g.~$\bm{q}(0)$, ${\bm{\dot{q}}}(0)$,
$\{\varepsilon_\nu (0) = 0\}$, $\{\dot{\varepsilon}_\nu (0)=0\}$.
\end{itemize}

\section{Existence Proof of Intraband Discrete Breathers}

In the previous section we have argued that the solution
$\bm{q}(t)$ of eq.(\ref{effenp}) (or equivalently
eq.(\ref{effenp1})) may converge towards an asymptotic solution
which might be periodic, quasiperiodic or weakly chaotic.

We prove now that indeed there may exist exact periodic solutions
under some conditions. When the phonon spectrum has an absolutely
continuous component, there may exist exact time periodic
solutions providing their frequency and higher harmonics are
outside the absolutely continuous spectrum $\mathcal{S}_{ac}$.
When there is a discrete component, those solutions may exist with
frequency and higher harmonics   inside the discrete spectrum
$\mathcal{S}_d$ but \textit{sufficiently} different from the
phonon frequencies $\omega_\nu$. In the present section, we prove
the existence of such periodic solutions.

Considering a time periodic solution of the uncoupled
($\lambda=0$) Hamiltonian $\mathcal{H}_0(\mathbf{p},\mathbf{q})$
with frequency $\Omega=2\pi/T$ and period $T$, we ask whether this
solution persists as an exact time periodic solution at the same
frequency when the coupling $\lambda$ is turned on at least up to
a nonvanishing value. $\Omega$ will be considered as a given
number which has to be chosen in an appropriate set in order this
solution persists.

We also need to assume some conditions concerning the time
periodic solutions which are generally fulfilled  in order to
answer this question. This time periodic solutions should belong
to a one parameter family of time periodic solutions. Thus, this
family can be represented locally in action angle variables. Its
energy $H_b(I)$  only depends on its action $I$  (at least defined
in some interval), then $H_b ^{\prime}(I)$ is the frequency of the
solution. We need to assume that this solution is not marginal at
the considered frequency $\Omega$ that is $H_b ^{\prime\prime}(I)
\neq 0$ when $H_b^{\prime}(I)=\Omega$. The possible existence of
such time periodic solutions in complex systems is already known
and for example may be based on earlier proofs of (interband) DBs
(for which the fundamental frequency and its harmonics are not
resonant with the phonon band) considered in \cite{MA94}.

The equation of motion for a periodic solution $\bm{q}(t; \Omega)$
with period $T=2\pi/\Omega$ of the anharmonic subsystem coupled to
the collection of harmonic oscillators is (see appendix
\ref{AppD})
\begin{equation}
\mathbf{\ddot{q}}(t;\Omega)+V'(\mathbf{q(t;\Omega)})
-\lambda^2\int\limits_0^T
\mathbf{G}(t-\tau;\Omega)\mathbf{q}(\tau;\Omega) d\tau =0
\label{effeq}
\end{equation}
with
\begin{equation}
 G_{\alpha,\beta}(t;\Omega) = \sum_{\nu}
 \frac{C_{\nu,\alpha}C_{\nu,\beta}}{2 \omega_{\nu} \sin (\omega_{\nu}\frac{T}{2})}
 \cos \omega_{\nu}(t-\frac{T}{2}), \,\, 0\leq t \leq T
 \label{kernelp}
 \end{equation}

which should no be confused with $G_{\alpha, \beta} (t)$.
$G_{\alpha \beta} (t; \Omega)$ is periodically continued for $t<0$
and $t > T$. A necessary condition for the existence of IDB with
frequency $\Omega$ is that the periodic kernel $G_{\alpha, \beta}
(t; \Omega)$ is a bounded and smooth function in $t$ for given
$\Omega$. This condition will be discussed in the following.

\subsection{Convergence of Kernel Contribution due to the Absolutely Continuous Spectrum}

According to the hypothesis, eqs. \eqref{abcspt}, \eqref{abcspt1},
the absolutely continuous part becomes the integral of a smooth
function of $\omega$ in the limit of $N$ to infinity:

\begin{eqnarray}
G_{\alpha,\beta}^{(ac)}(t,\Omega)&=& \sum_{\omega_{\nu} \in
\mathcal{S}_{ac}} \frac{C_{\nu,\alpha}C_{\nu,\beta}}
{2\omega_{\nu} \sin \pi
  \frac{\omega_{\nu}}{\Omega}}
\cos \omega_{\nu} \left(t-\frac{T}{2}\right)\nonumber\\
&=& \int\limits_{\omega \in \mathcal{S}_{ac}}
 \frac{c_{\alpha}(\omega)c_{\beta}(\omega) }
{2\omega \sin \pi  \frac{\omega}{\Omega}} \cos \omega
\left(t-\frac{T}{2}\right) \mu(\omega)\, d\omega \label{kernec}
\end{eqnarray}

which diverges for $\omega=p\Omega$ if $p\Omega \in
\mathcal{S}_{ac}$ for some nonzero integer $p$. Then,
$\mathbf{G}^{(ac)}(t; \Omega)$ is not  a smooth bounded function
of time. Note, the divergence at $\omega=0$ ($p=0$) is harmless
because of the hypothesis eq.\eqref{hypf1} of finite
reorganization energy. Thus, the contribution of the absolutely
continuous spectrum, $\mathbf{G}^{(ac)}(t; \Omega)$, is a smooth
bounded function of time when

\begin{equation}
\Omega \notin \bigcup_{p>0} \frac{\mathcal{S}_{ac}}{p}
\label{cond1} \quad .
\end{equation}

This is the standard condition expected from earlier theories for
the existence of DBs which assume the absence of resonance of the
fundamental frequency and its higher harmonics with the absolutely
(continuous) phonon spectrum. It is consistent with the
conclusions drawn from eq.\eqref{freqd} in (i) of subsection 4.2.

\subsection{Convergence of the Kernel Contribution due to
the Discrete Spectrum}

The contribution due to the discrete spectrum

\begin{equation} G_{\alpha,\beta}^{(d)}(t;\Omega)=
\sum_{\omega_{\nu} \in \mathcal{S}_d}
\frac{C_{\nu,\alpha}C_{\nu,\beta}} {2\omega_{\nu} \sin \pi
\frac{\omega_{\nu}}{\Omega}}\cos \omega_{\nu}
\left(t-\frac{T}{2}\right) \label{kerned}
\end{equation}

cannot be transformed into the integral of a smooth function.
Eq.\eqref{kerned} clearly demonstrates the problem of small
denominators which makes the proof of the existence of IDB with
frequency $\Omega$ in presence of a dense discrete phonon spectrum
nontrivial. This problem of small denominator is quite similar to
that we encountered in (ii) of subsection 4.2. In the following we
will study this small denominator problem.

For given constant $c>0$ and given $\Omega$, we define a closed
interval or a finite union of closed intervals $\mathcal{I}_d$
strictly included in the discrete spectrum $\mathcal{S}_d$ such
that for $\omega \in \mathcal{I}_d$, $|\sin \pi
\frac{\omega}{\Omega}| \leq c.$ $c$ is arbitrary except that it
must be chosen small enough in order $\mathcal{I}_d$ does not
contain the edges of $\mathcal{S}_d$ (mobility edges). This set
$\mathcal{I}_d$ contains the values of $p\Omega \in
\mathcal{S}_d$. Then, for $\omega_{\nu} \in \mathcal{S}_d
\backslash  \mathcal{I}_d$, the denominators $|\sin \pi
\frac{\omega_\nu}{\Omega}| > c>0$  are bounded from below by $c$.

We split again the discrete kernel \eqref{kerned} into two parts
$\mathbf{G}^{(d)}(t;\Omega) =\mathbf{G}^{(dc)}(t;\Omega)+
\mathbf{G}^{(ds)}(t;\Omega)$ where

\begin{equation}
G_{\alpha,\beta}^{(dc)}(t;\Omega)= \sum_{\omega_{\nu} \in
\mathcal{S}_d \backslash  \mathcal{I}_d}
 \frac{C_{\nu,\alpha}C_{\nu,\beta}}
{2\omega_{\nu} \sin \pi \frac{\omega_{\nu}}{\Omega}}\cos
\omega_{\nu} \left(t-\frac{T}{2}\right) \quad . \label{kernedc}
\end{equation}

is a smooth bounded function of time because of the absence of
small denominators  (they are always larger than $c\omega_{\nu}$
and therefore the sum converges because of the hypothesis of
finite reorganization energy (\ref{hypf1}); the divergence at
$\omega_\nu=0$ is harmless since $C_{\nu, \alpha} \sim \omega_\nu$
for $\omega_\nu \neq 0$, as already mentioned before)

The small denominators are in the second part

\begin{equation}
G_{\alpha,\beta}^{(ds)}(t;\Omega)= \sum_{\omega_{\nu} \in
\mathcal{I}_d}
 \frac{C_{\nu,\alpha}C_{\nu,\beta}}
{2\omega_{\nu} \sin \pi \frac{\omega_{\nu}}{\Omega}}\cos
\omega_{\nu} \left(t-\frac{T}{2}\right) \quad .\label{kerneds}
\end{equation}

In order to insure the convergence of \eqref{kerneds}, we choose
$\Omega$ in  a particular subset $\mathcal{C}(k)$ of the real
positive axis defined  such that for  all $\omega_{\nu} \in
\mathcal{S}_d$, we have

\begin{equation}
\frac{|\mathbf{C}_{\nu}|^{2}}{2  \omega_{\nu} |\sin \pi
\frac{\omega_{\nu}}{\Omega}|} < k
\frac{|\mathbf{C}_{_\nu}|}{\omega_{\nu}}\quad . \label{ord3}
\end{equation}

Then, if $\Omega\in \mathcal{C}(k)$, we have

\begin{equation}
|\mathbf{G}^{(ds)}(t;\Omega)|= \sum_{\omega_{\nu} \in
\mathcal{I}_d}
 \frac{|\mathbf{C}_{\nu}|^2}
{2\omega_{\nu} |\sin \pi \frac{\omega_{\nu}}{\Omega}|} < k
\sum_{\omega_{\nu} \in \mathcal{I}_d \subset \mathcal{S}_d}
\frac{|\mathbf{C}_{_\nu}|}{\omega_{\nu}}= kB(\mathcal{I}_d)
\label{boundk}
\end{equation}

which is convergent according to the hypothesis \eqref{hypf2}.
Consequently, for $\Omega \in \mathcal{C}(k)$, the discrete
spectrum contribution $\mathbf{G}^{(d)}(t; \Omega)$ to the kernel
$\mathbf{G} (t;\Omega)$ in eq. \eqref{kerned} is a smooth bounded
function of time.

We now have to prove that $\mathcal{C}(k)$ is not an empty set if
we choose the parameter $k$ in an appropriate way. First, we
choose  $k$  large enough in order $0<\epsilon_{\nu}<\frac{1}{2}$
can be defined (uniquely) for all $\omega_{\nu}\in \mathcal{S}_d$
by
\begin{equation}
   \frac{1}{2k} |\mathbf{C}_{\nu}|  = \sin \pi \epsilon_{\nu}
   \quad . \label{gap}
\end{equation}

Then, condition \eqref{ord3} becomes

\begin{equation}
   \sin \pi \epsilon_{\nu}  \leq |\sin \pi
    \frac{\omega_{\nu}}{\Omega}|
    \label{ord2}
\end{equation}

which requires for all $p>0$ and all $\omega_{\nu} \in
\mathcal{S}_d$
\begin{equation}
\Omega \notin \mathcal{D}(k) =\bigcup_{p,\omega_{\nu} \in
\mathcal{S}_d}
\left]\frac{\omega_{\nu}}{p+\epsilon_{\nu}},\frac{\omega_{\nu}}{p-\epsilon_{\nu}}\right[
\quad . \label{nuset}
\end{equation}
$\mathcal{C}(k)$ is the complementary set
 $\mathcal{C}(k)=\mathcal{R}_+ \backslash \mathcal{D}(k)$. The total measure of the gaps of $\mathcal{C}(k)$ can be easily
bounded from definition (\ref{nuset}) for $\omega_{\nu} \in
\mathcal{I}_d$ where $\mathcal{I}_d$ is any interval or union of
intervals strictly included in $\mathcal{S}_d$  as
\begin{equation}
\mu( \mathcal{D}(k)) < \sum_{\omega_{\nu} \in \mathcal{I}_d}
\sum_{p>0}
\frac{2\epsilon_{\nu}\omega_{\nu}}{p^2-\epsilon_{\nu}^2} <
\sum_{\omega_{\nu} \in \mathcal{S}_d}2\epsilon_{\nu}\omega_{c}
 \left(\sum_{p>0} \frac{1}{p^2-1/4}\right) \quad .
\label{gapmeas}
\end{equation}

Since $0<\epsilon_{\nu} <\frac{1}{2}$, we have $0<2 \epsilon_{\nu}
<\sin{ \pi \epsilon_\nu}= \frac{1}{2k} |\mathbf{C}_{\nu}| <
\frac{\omega_c}{2k} \frac{|\mathbf{C}_{\nu}|}{\omega_{\nu}} $
,using definition \eqref{gap} and inequality \eqref{ord3}  . Then
using condition \eqref{hypf2}, we obtain

\begin{equation}
\mu(\mathcal{D}(k)) < \frac{K}{k}B(\mathcal{I}_d) \quad .
\label{gapmeas2}
\end{equation}

where
$$K=\frac{\omega_c^2}{2} \sum_{p>0} \frac{1}{p^2-1/4} <+\infty$$

Choosing $k$ large, the Lebesgue measure of  $\mathcal{D}(k)$
becomes small so that the measure of the Cantor set
$\mathcal{C}(k)$ tends to full measure in $ \bigcup_{p>0}
\frac{\mathcal{I}_d}{p}$.

\subsection{Proof of existence of IDB}

After having proved the existence of a set $\mathcal{C}(k)$ s.t.
$G_{\alpha, \beta} (t; \Omega)$ is a bounded and smooth function
of $\Omega \in \mathcal{C}(k)$ we now apply the implicit function
theorem to prove the existence of IDB.

We consider the Banach space $\mathcal{L}_1$ of T-periodic
differentiable functions $(\bm{p}(t),\bm{q}(t))$ with continuous
time derivatives which are time periodic
$\bm{p}(t+T)=\bm{p}(t),\bm{q}(t+T)=\bm{q}(t)$  and fulfill time
reversibility: $\bm{p}(t)=-\bm{p}(-t)$ and $\bm{q}(t)=\bm{q}(-t)$.
The norm in  $\mathcal{L}_1$ is defined as
$||\bm{p}(t),\bm{q}(t)||_1= \sup_t\Bigl\{|\bm{p}(t)|,|\bm{q}(t)|,$\\
$|\bm{\dot{p}}(t)|,|\bm{\dot{q}}(t)|\Bigr\}$. We consider the
Banach space $\mathcal{L}_0$ of T-periodic continuous and time
reversible functions $(\bm{v}(t),\bm{w}(t))$ with the norm
$||\bm{v}(t),\bm{w}(t)||_0=
\sup_t\Bigl\{|\bm{v}(t)|,|\bm{w}(t)|\Bigr\}$.

Let us consider for a given $\Omega \in \mathcal{C}(k)$, the map
$\mathcal{T}(\bm{p},\bm{q};\lambda)$  defined as
\begin{equation}
\left(\begin{array}{c} \bm{v} \\ \bm{w}
\end{array}\right)=\mathcal{T}(\bm{p},\bm{q};\lambda)=
\left(\begin{array}{c} \bm{p}-\bm{\dot{q}} \\
\bm{\dot{p}}+V^{\prime}(\bm{q}) -\lambda^2\int_0^T
\bm{G}(t-\tau;\Omega) \bm{q}(\tau) d\tau
\end{array} \right)
\end{equation}

It maps $\mathcal{L}_1$ in  $\mathcal{L}_0$ and is continuous and
differentiable with respect to $(\bm{p},\bm{q})$ and $\lambda$.
The zeros of $\mathcal{T}(\bm{p},\bm{q};\lambda)$ yield time
periodic solutions of eq. \eqref{effeq}.

At $\lambda=0$, let us assume that the anharmonic subsystem has a
periodic and time reversible solution $\bm{q_0}(t)$
 with period $T=2\pi/\Omega$ with $\Omega \in \mathcal{C}(k)$.
If $\partial \mathcal{T}(\bm{p},\bm{q};0) $ is invertible at
$(\bm{p_0},\bm{q_0})$ then this solution can be continued in
$\lambda$ in a neighborhood of zero up to a nonzero value. We
have

\begin{equation}
\partial \mathcal{T}(\bm{p_0},\bm{q_0};0)
\left(\begin{array}{c} \delta \bm{p} \\ \delta \bm{q}
\end{array}\right)= \left(\begin{array}{c} \delta \bm{p}-\delta
\bm{\dot{q}} \\ \delta \bm{\dot{p}}+V^{\prime\prime}(\bm{q_0})
\delta \bm{q} \end{array}\right)
\label{tangentmap}
\end{equation}

The condition for invertibality of $\partial \mathcal{T}(\bm{p_0},\bm{q_0};0) $  follows from the
Floquet analysis \cite{Aub97}. Setting the r.h.s. of \eqref{tangentmap} to zero yields the linearized equation of motion
with respect to the periodic and time reversible solution $(\bm{p_0}(t),\bm{q_0}(t))$
of the anharmonic subsystem. Integration with respect to time over one period leads to the Floquet matrix
of the unperturbed DB. Then, $\partial \mathcal{T}(\bm{p_0},\bm{q_0};0) $ is invertible, if the Floquet matrix
has only a single eigenvalue which equals one. Since its corresponding eigenfunction is antisymmetric with
respect to time reversion, this eigenmode is harmless.

\section{Summary and conclusions}

We have assumed that a general macroscopic classical particle
system with arbitrary interactions can be decomposed into a
microscopic or macroscopic anharmonic subsystem with coordinates,
$\bm{q}= (q_{\alpha})$, and anharmonic interaction energy
$V(\bm{q})$, and a remaining macroscopic part with harmonic
interactions. In order that the linear spectrum may possess a
continuous part one has to take the thermodynamic limit of the
harmonic subsystem. The interaction between the harmonic
displacement variables $u_i$ at site $i$ with $q_{\alpha}$ is
assumed to be bilinear with coupling constants $\lambda
C_{i,\alpha}$. $\lambda$ is introduced for technical convenience.
A realistic system for which these assumptions may apply is, e.g.
a complex impurity embedded in a lattice where the latter can be
treated in harmonic approximation. The harmonic part can be
diagonalized by introducing normal mode coordinates $u_\nu$ for
eigenmodes with frequency $\omega_\nu$. Then our model takes the
form which we call a Zwanzig-Kac model \cite {3,4} for which the
harmonic degrees of freedom can be eliminated. This results in the
integro-differential equation eq. \eqref{effenp} for $\bm{q}(t)$,
which is of Langevin type with memory kernels $\bm{G}(t)$. The
properties of $\bm{G}(t)$, or equivalently of the dissipation
kernels $\bm{\Gamma}(t)= - {\bm{\dot{G}}} (t)$, depend strongly on
the spectrum of the harmonic part.

The elimination of a
macroscopic number of degrees of freedom leads to fluctuating
forces and (as always believed) to dissipation. The former vanish
at zero temperature. Then it is the common belief that $\bm{q}(t)$
for arbitrary initial conditions $(\bm{q}(0)$, $\bm{\dot{q}}(0))$
converges to one of the local minima of the effective potential
$V_{\rm eff}(\bm{q})$. However, this is only true if the phonon
spectrum of the harmonic part is absolutely continuous with
infinite support. In general, its support may be not infinite and
in addition  there may also exist a discrete spectrum, e.g. due to
disorder. It has been one of the major goals of our contribution
to prove that energy dissipation strongly depends on both, the
specific properties of the phonon spectrum and of the trajectories
$\bm{q}(t)$. For rather general conditions we have succeeded to derive an exact formula for the
energy dissipation rate ${\overline{\dot{E}}}_T$ of the anharmonic
system averaged over a time interval of length $T$. For $T
\rightarrow \infty$ it only involves $\tilde{\Gamma}_{\alpha,
\beta} (\omega) \tilde{K}_{\beta, \alpha} (\omega)$, the product
of the Fourier transform of the dissipation kernels
$\Gamma_{\alpha, \beta} (t)$ and of the velocity-velocity
correlation functions $K_{\alpha, \beta} (t)$. For a finite
initial energy and a Hamiltonian bounded from below it must be
${\overline{\dot{E}}}_{T=\infty}=0$. This is possible, only, if
either $\tilde{K}_{\alpha, \beta} (\omega)\equiv0$, which holds for $\bm{q}(t)
\rightarrow \bm{q}_*$ (local minimum of $V_{\rm eff}(\bm{q}))$ for
$t \rightarrow \infty$, or the support of the measures
$\bm{\tilde{\Gamma}} (\omega) d (\omega)$ and ${\bm{\tilde{K}}}
(\omega) d \omega$ have no overlap. Besides the convergence of
$\bm{q}(t)$ to one of the local minimum $\bm{q}_*$ of $V_{\rm eff}
(\bm{q})$, the nonoverlapping  criteria allows {\it possible}
asymptotic solutions which are periodic, quasiperiodic (or almost
periodic) or may be even weakly chaotic.

If weakly chaotic
asymptotic trajectories really exist, the corresponding measure
${\bm{\tilde{K}}} (\omega) d \omega$ must be singular continuous.
An absolutely continuous measure part of
${\bm{\tilde{K}}} (\omega) d \omega$ is  ruled
out, since due to a theorem (proved in appendix C) its support
must be the full real line, which always has an overlap with the
support of ${\bm{\tilde{\Gamma}}}(\omega) d \omega$.

Quasiperiodic (or almost periodic) aymptotic solutions  can not
exist as well, if ${\bm{\tilde{\Gamma}}}_{ac} (\omega) d \omega$
has an absolutely continuous  component .  Because, with
$\Omega_1$ and $\Omega_2$ in the support of ${\bm{\tilde{K}}}
(\omega) d \omega$ also $n_1 \Omega_1 + n_2 \Omega_2$ with $n_i$
integers will be in that support. Then there exist an infinite
number of integer pairs ($n_1$, $n_2$) , such that $n_1 \Omega_1 +
n_2 \Omega_2$ has an overlap with  the support of
${\bm{\tilde{\Gamma}}}_{ac} (\omega) d \omega$, provided
$\Omega_1/\Omega_2$ is an irrational number. Quasiperiodic (or
almost periodic) solutions may exist if
${\bm{\tilde{\Gamma}}}(\omega) d \omega$ has a discrete support,
only. We do not have a proof for this. For a special class of
models with disorder the existence of such solutions were proven
rigorously \cite{FSW}. The model studied in the present paper, may
posses almost periodic solutions, as well.

Periodic asymptotic solutions can occur when
${\bm{\tilde{\Gamma}}}_{ac} (\omega) d \omega$ has a {\it compact}
absolutely  continuous, a singular continuous and/or a discrete component.
Their existence is supported by the calculation
of the dissipation rate ${\overline{\dot{E}}}_T$ for a sinusoidal
force with frequency $\Omega$. In that case a zero measure set
$\mathcal{L}$ exists (see appendix B), such that $|E_T|$
is bounded for all $\Omega \in \mathcal{R} \backslash
\mathcal{L}$, the full measure complement of $\mathcal{L}$. In
that case a Fourier component of $\bm{q}(t)$ with frequency
$\Omega$ would be undamped. That  dissipation might be even more
subtle has been demonstrated by the proof of the existence of a
subset $\mathcal{L}'$ of $\mathcal{L}$ (see appendix B). The zero measure set
$\mathcal{L}'$ does not contain any of the phonon frequencies
$\omega_\nu$ with probability one (assuming $\omega_\nu$ and
$\omega_{\nu'}$, $\nu \neq \nu'$ as uncorrelated). Although
$\Omega \in \mathcal{L}'$ is not in resonance with any phonon
frequency there is dissipation. Depending on $\Omega$ there may
exist subdissipative and superdissipative behavior, i.e.~$E_T$
increases, respectively, slower and faster than $T$, where the
linear $T$-dependence is the behavior of ordinary dissipation.
Whether $E_T$ follows a power law $T^\alpha$ with $0 < \alpha <1 $
(subdissipative) and $1 < \alpha \leq 2$ (superdissipative), as
conjectured in (ii) of subsection 4.2, is not obvious. Anyway, we
have shown that dissipation can be rather complex.\\

The existence of nonrelaxing solutions also shows that the phonon
bath can act as an anomalous thermostat. Depending on the initial
conditions for $\bm{q}(t)$, the perturbed anharmonic system may
not relax to equilibrium, provided the initial energy is finite,
i.e.~of order $N^0$ in the total number of d.o.f. There may even
exist a sharp transition between relaxing and nonrelaxing
situations. As shown recently \cite{29} for a reduced Fermi-Pasta-Ulam
model with a single anharmonic bond embedded in an infinite
harmonic chain there exists a critical excitation amplitude $A_c$
of the anharmonic bond such that the initial configuration relaxes
to equilibrium for $A<A_c$ and to a discrete breather for $A >
A_c$. That   such an incomplete energy spreading can occur has
also been proven for a general discrete nonlinear Schr\"odinger
equation in any spatial dimension and without and with disorder
\cite{28}.


Another important goal of our paper has been to prove the
existence of IDBs. The existence of IDBs with frequency $\Omega$
has been shown by proving the smoothness and boundedness of the
{\it periodic} kernel $\bm{G}(t;\Omega)$ with respect to time $t$.
In that case, following reference \cite{MA94}, the implicit
function theorem can be applied to the time periodic solutions at
frequency $\Omega$ of the uncoupled subsystem, i.e. at $\lambda =
0$, in order to prove the existence of time periodic solutions at
the same frequency $\Omega$ for a nonzero value of $\lambda$. The
boundedness of $\bm{G}(t; \Omega)$ is not obvious due to the
occurrence of small denominators (cf. eq. \eqref{kernelp}). These
small denominators make more explicit the resonance problem between the
frequency $\Omega$ and the harmonic mode frequencies $\omega_\nu$
(which is also the basic problem, discussed in Ref. \cite{AF88,AF91,AFS88}).
It is quite analogous to the small denominator
problem in secular perturbation theory in classical mechanics. We
have proven that there exists a Cantor set $\mathcal{C}(k)$ of
finite Lebesgue measure such that $\bm{G}(t;\Omega)$ is smooth and
bounded for all $\Omega\in\mathcal{C}(k)$. This implies that
the periodic solutions of  the isolated anharmonic
subsystem with frequency  $\Omega\in\mathcal{C}(k)$ can be
continued  up to a small but finite interaction
strength between the anharmonic and harmonic subsystem. Therefore, there
are IDBs for all $\Omega\in\mathcal{C}(k)$, provided $\lambda$ is
small enough. The approach in Ref. \cite{AF88,AF91,AFS88}
leading to the same conclusion for a different class of models, is partly
complementary to ours. The authors of  Ref. \cite{AF88,AF91,AFS88} consider
a fully anharmonic system with quartic anharmonicity whereas our model consists of an anharmonic subsystem embedded in a harmonic lattice.
The anharmonicity can be rather general.
For our class of model it is the elimination of the harmonic d.o.f.
which finally leads  to the small denominators in its explicit form of eq. \eqref{kernelp}.

There are several open problems which could be studied but which
go beyond the present contribution. \textit{First}, the linear stability of the
IBDs is one of them. \textit{Second}, the existence of quasiperiodic or
almost periodic solutions, which have already been proven for a special model \cite{FSW}, and particularly of weakly chaotic solutions of
the present model could be investigated. \textit{Third}, there is the
question whether an arbitrary initial condition $(\bm{q}(0)$,
$\bm{\dot{q}}(0))$ with large enough energy does relax completely
or not. If it does not, where does it converge to? Since
almost periodic solutions (which correspond to the existence of
KAM tori) should be linearly stable, i.e. all Lyapunov exponents should be zero,
one should exclude convergence to \textit{true} tori. Thus, the limit solution could only be at the ``edge`` of the KAM tori region with still discrete or perhaps singular continuous spectrum?
\textit{Fourth}, if both scenarios, i.e. complete and incomplete spreading
(depending on the initial condition) exist, it will be interesting to study the existence of a sharp transition between both.

{\bf Acknowledgement:} This work has been started at the Johannes
Gutenberg University in Mainz (Germany). One of us (S.A.) thanks
for the financial support. It has been extended and completed at the
Max-Planck-Institut f\"ur Physik komplexer Systeme in Dresden
(Germany) when both authors were visitors for a longer period of
time. We gratefully acknowledge the MPIPKS for financial support
and its hospitality. We also would like to thank J\"urg Fr\"ohlich
for his critical and helpful comments on this manuscript.

\appendix
\newpage
\section{Relation between  the time Fourier transform of a bounded function
and the Fourier transform of its autocorrelation function}

\label{AppA} We prove that (if defined) the limit of the square
modulus of the Fourier transform (eq.(\ref{velfor})) of
$g_{\alpha; T}(t)$ (eq.(\ref{veldef})) is identical to the Fourier
transform of the matrix of autocorrelation  functions of the
velocities, but in the measure sense only. Since these Fourier
transforms might not be smooth functions,  they should be
considered as positive measures which may have a discrete or
singular continuous part. In that case a direct analytical
treatment would lead to convergence problems because the kernel of
the involved integrals, may not vanish at infinity. It is thus
worthwhile to present a more rigorous treatment showing a relation
not in terms of smooth functions, but of measure. For simplicity,
we consider here only the case with a single component for the
velocity vector but the proof would readily extend to cases with
arbitrarily many components.

Let us consider a smooth bounded function $q(t)$ (which could be
for example the time dependant velocity we consider above) and its
Fourier transform over a finite time $T$ weighted by $1/\sqrt{T}$
$$\tilde{g}_T(\omega) = \frac{1}{\sqrt{T}} \int_0^T q(t) e^{i\omega t} dt .$$
This  is a smooth function of $\omega$ for any finite $T$. Its
square modulus  defines an absolutely continuous positive measure
$d\sigma_T(\omega)=|\tilde{g}_T(\omega)|^2 d\omega$. We must now
assume that $\lim_{T\rightarrow +\infty} \sigma_T(\omega) =
\sigma(\omega)$ is defined in the sense of   weak convergence of
these measures  \cite{mathtxt}. The limit is a priori any kind of
measure absolutely continuous, singular continuous or discrete.
This assumption means by definition that for any bounded
continuous  function $f(\omega)$  (for the uniform topology)  we
have $\lim_{T\rightarrow +\infty} \int f(\omega)
d\sigma_T(\omega)=\int f(\omega) d\sigma(\omega)$.

Let us consider the subset of bounded continuous  functions
$f(\omega)$ with a compact support for ensuring the existence of a
smooth Fourier transform $F(t)=  \int f(\omega)  e^{-i\omega t}
d\omega $ which fulfills  the condition $ \int |F(t)| dt< +
\infty$ and $\,\int |f(\omega)| d\omega < +\infty$. Then,
$$\int f(\omega) d\sigma_T(\omega)=  \int  f(\omega)
\left(\frac{1}{T}  \int_0^T \int_0^T q(t) q(t^{\prime})
e^{-i\omega( t-t^{\prime})} dt dt^{\prime} \right) d\omega$$
$$=\frac{1}{T}  \int_0^T \int_0^T q(t) q(t^{\prime})  \left( \int
 f(\omega)  e^{-i\omega( t-t^{\prime})} d\omega \right)  dt dt^{\prime}
 =\frac{1}{T}  \int_0^T \int_0^T q(t) q(t^{\prime}) F(t-t^{\prime})  dt dt^{\prime}.$$
Since $|q(t)|$ is bounded, this double integral is absolutely
convergent in the limit $T\rightarrow +\infty$,  it readily comes
out  with the  new integration variables $t$ and
$\tau=t-t^{\prime}$ that for any $L_1$ smooth function $F(t)$, we
have
\begin{equation}
\lim_{T\rightarrow +\infty} \int f(\omega) d\sigma_T(\omega)= \int
f(\omega) d\sigma(\omega)= \int_{-\infty}^{+\infty} F(\tau)
\left(\lim_{T\rightarrow +\infty} \frac{1}{T}  \int_0^T  q(t)
q(t+\tau)  dt\right)  d\tau \label{appa2}
\end{equation}
Consequently, for most values of $\tau$, the autocorrelation
function
\begin{equation}
C(\tau)=C(-\tau) =\lim_{T\rightarrow +\infty} \frac{1}{T}
\int_0^T  q(t) q(t+\tau)  dt \label{autocorapp}
\end{equation}
is a well-defined function (since  $|C(\tau)|$ is obviously
bounded). Defining, the Fourier transform measure $dc(\omega)=
(\int C(\tau) e^{i \omega \tau} d\tau) d\omega$, it comes out
$$ \int f(\omega) d\sigma(\omega)= \int f(\omega) dc(\omega) $$
for all continuous bounded functions $f(\omega)$ with compact
support. We can now take sequences of continuous function with
compact support converging toward arbitrary continuous functions
without compact support which proves the equality remains
fulfilled for arbitrary continuous functions $f(\omega)$ Thus, we
prove the equality of the Fourier transform of the autocorrelation
function $dc(\omega)$ considered as a measure with the limit at
infinite time of the modulus square of finite time Fourier
transform of the trajectory   averaged over time
$d\sigma(\omega)$.

It appears from eq.\ref{appa2} that assuming $d\sigma_T(\omega)$
has a limit measure for $T\rightarrow +\infty$ or that the
autocorrelation function (\ref{autocorapp}) is defined are
equivalent conditions.

\section{Derivation of set $\mathcal{L}$ and $\mathcal{L}'$}

A sufficient condition for the absolute convergence of series
\eqref{freqd4} is the convergence of

\begin{equation}
\mathcal{A} (\Omega) =\sum\limits_\nu a_\nu
\Big(\frac{1}{(\omega_\nu + \Omega)^2} + \frac{1}{(\omega_\nu-
\Omega)^2} \Big) . \label{B.1}
\end{equation}

Given an arbitrary convergent positive series $\{s_{\nu}\}$ with
$s_{\nu}>0$ which decays slower than $a_{\nu}$ for example as a
power law of $\nu$ such as $s_{\nu} = \frac{1}{\nu^{1+\epsilon}}$
(with $\epsilon>0$) and given some positive integer $k$, we can
make that serie $\mathcal{A}(\Omega)$  convergent if we choose
$\Omega$ such that $|\Omega+\omega_{\nu}| \geq r_{\nu}/k$ and
$|\Omega-\omega_{\nu}| \geq r_{\nu}/k$ with
$r_{\nu}=(\frac{a_{\nu}}{s_{\nu}})^{1/2}$. This condition means
$\Omega\notin \mathcal{L}_k$ where $\mathcal{L}_k$ is  the union
of open intervals

\begin{equation}
\Omega \notin \mathcal{L}_k = \left(  \bigcup_{\nu}~ \left]
\omega_{\nu}-\frac{r_{\nu}}{k},\omega_{\nu}+\frac{r_{\nu}}{k}
\right[ \right) \bigcup \left(  \bigcup_{\nu} \left]
-\omega_{\nu}-\frac{r_{\nu}}{k},-\omega_{\nu}+\frac{r_{\nu}}{k}
\right] \right) \label{setl}
\end{equation}

The Lebesgue measure of this union of intervals is bounded as
$$\mu( \mathcal{L}_k) \leq \frac{1}{k} \sum_{\nu} 4 r_{\nu}= 4 \frac{S}{k}$$  Since $a_{\nu}$
is supposed to converge exponentially to zero and since $s_{\nu}$
has been chosen to go to zero as a power law, series $S=
\sum_{\nu} r_{\nu}= (\frac{a_{\nu}}{s_{\nu}})^{1/2} < \infty$ is
absolutely  convergent for $\Omega$ chosen in the complementary
set  of $\mathcal{L}$ \footnote{Note that   $a_{\nu}$ does not
have necessarily to decay exponentially but could also decay as  a
power law with an exponent $\beta$ strictly larger than $3$. This
is sufficient for having $\sum_{\nu} r_{\nu} <+\infty$. Indeed if
we assume $0<a_{\nu}< \frac{K}{\nu^{\beta}}$, then $r_{\nu} <
\frac{K^{1/2}}{\nu^{(\beta-1-\epsilon)/2}}$ so that if $\beta>3$
this series can be made convergent with an appropriate choice of
$\epsilon>0$.}. Consequently, we have $\mu( \mathcal{L}_k) \leq
4S/k$.

$\mathcal{L}_k$ is an open set since it is the union of open sets
. It is dense on the discrete spectrum $\mathcal{S}_d$ since it
contains $\omega_{\nu}$ for all $\nu$ which are dense in
$\mathcal{S}_d$. We have $\mathcal{L}_{k+1} \subset
\mathcal{L}_k$. The set  $\mathcal{L} = \bigcap_k \mathcal{L}_k$
which is a countable intersection of open sets $\mathcal{L}_k$
dense on $\mathcal{S}_d$,  is not an open set but is  also dense
in $\mathcal{S}_d$ (according to Baire theorem \cite{mathtxt}). When
$\Omega \notin \mathcal{L}$, there exists $k$ such that $\Omega
\notin \mathcal{L}_k$, series (\ref{freqd4})  is absolutely
convergent  and remains  bounded at infinite time Thus, the energy
transferred to the phonon bath remains
 finite and bounded at all time. This  set $\mathcal{L}$ has zero Lebesgue
measure since $\mu( \mathcal{L}) \leq \mu( \mathcal{L}_k) \leq
4S/k$ for $k$ arbitrarily large. Thus, the  complementary set
$\mathcal{R} \backslash \mathcal{L}$ has full Lebesgue measure.

By analogy with the theory of Liouville numbers, we now prove that
this  set $\mathcal{L}$ contains a (zero measure) subset
$\mathcal{L}^{\prime} \subset \mathcal{L}$  for which series
(\ref{freqd4}) surely diverges so that there is (anomalous) energy
dissipation. We can also prove that this set
$\mathcal{L}^{\prime}$ does not contain any $\omega_{\nu}$  with
probability one if we assume  the distribution of eigenvalues
$\omega_{\nu}$ is uniform enough on $\mathcal{S}_d$. Actually, the
probability distribution of these eigenvalues is nothing but the
measure density of states $g(\omega) d\omega$. We have to assume
that this measure is absolutely continuous and moreover that the
distribution probability of a pair of two eigenvalues
$\omega_{\nu}$ and $\omega_{\nu^{\prime}}$ for $\nu \neq
\nu^{\prime}$  is  obtained as the  product of probability
distribution of single eigenvalues (i.e. the eigenvalues are
uncorrelated).

The energy transferred to the phonon bath given by series
(\ref{freqd4}) diverges  if its sequence of coefficients  does not
go to zero. This situation  surely occurs when for any given
$\Omega$ there is an infinite subsequence $\omega_{\nu_n}$ such
that $|\Omega-\omega_{\nu_n}| < k a_{\nu_n}^{1/2}$ where $k$ is
some positive constant. This condition is equivalent to say
$\Omega \in \mathcal{L}^{\prime}$ where $\mathcal{L}^{\prime}=
\bigcap_N \mathcal{L}^{\prime}_N$ and $\mathcal{L}^{\prime}_N$ is
the union of open intervals

\begin{equation}
\mathcal{L}^{\prime}_N = \left(\bigcup_{\nu \geq N} \left]
\omega_{\nu} - k a_{\nu}^{1/2}, \omega_{\nu} + k
a_{\nu}^{1/2}\right[ \right)\cup \left(\bigcup_{\nu \geq N} \left]
-\omega_{\nu} - k a_{\nu}^{1/2}, -\omega_{\nu} + k
a_{\nu}^{1/2}\right[ \right) \label{setlp}
\end{equation}

$\mathcal{L}^{\prime}_N$  is a dense open set on $\mathcal{S}_d$
and $\mathcal{L}^{\prime}_{N+1} \subset  \mathcal{L}^{\prime}_N$
for all  $N$. Thus the countable intersection
$\mathcal{L}^{\prime}= \bigcap_N \mathcal{L}^{\prime}_N$  is a
dense set  (according again to Baire theorem \cite{mathtxt}). The
Lebesgue measure $\mu(\mathcal{L}^{\prime})$ of this set
$\mathcal{L}^{\prime}$ is zero because $\mu(\mathcal{L}^{\prime})
< \mu(\mathcal{L}_N^{\prime}) < 4k \sum_{\nu\geq N}
(a_{\nu})^{1/2}$ which is the rest of an absolutely convergent
series since  $a_{\nu}$ decays faster than $1/\nu^{2+\epsilon}$,
since we assumed an exponential decay.

We have now to prove that  giving an arbitrary $\omega_{\nu}$, it
could belong to this set $\mathcal{L}^{\prime}$  only with
probability $0$. If $\omega_{\nu} \in \mathcal{L}^{\prime}$ we
have $\omega_{\nu} \in \mathcal{L}^{\prime}_N$, for any $N$. Let
us call $P$ the probability for $\omega_{\nu}$ to belong to
$\mathcal{L}^{\prime} = \bigcap \mathcal{L}^{\prime}_N$. and let
us call $P_N$, the probability that $\omega_{\nu}$  belongs to the
unions of intervals $\mathcal{L}^{\prime}_N$. We have $0\leq P\leq
P_N$. We choose $N>\nu$ in order $\omega_{\nu}$ is not the center
of an interval of $\mathcal{L}^{\prime}_N$. Then, since we assumed
there is no correlation between the probability distribution of
$\omega_{\nu^{\prime}}$  with that of $\omega_{\nu}$ the
probability for $\omega_{\nu}$ to belong to
$\mathcal{L}^{\prime}_N$  is   $P_N=\int_{\mathcal{L}^{\prime}_N}
g(\omega) d \omega$ where $g(\omega)d\omega$ is the density of
states. When $N$ goes to infinity, the Lebesgue measure of
$\mathcal{L}^{\prime}_N$ goes to zero. Consequently $P\leq P_N
~\forall N$ has to be zero. Since the set of $\omega_{\nu}$ is
countable, there is no $\omega_{\nu} \in \mathcal{L}^{\prime}$
with probability one. \footnote{This proof should be extendable
when the probability correlations between $\omega_{\nu}$ and
$\omega_{\nu^{\prime}}$   decay fast enough as
$|\nu-\nu^{\prime}|$ diverges.}.

\newpage

\section{Lemma and its Proof}
\label{AppC}

\textbf{Lemma} \textit{Let us consider a measurable subset
$\mathcal{E} \subseteq \mathcal{R}$ of the real line with the
following property:}

\begin{eqnarray}
&&\forall \omega_1\in \mathcal{E}~ \mbox{and} ~ \forall \omega_2\in \mathcal{E}\nonumber \\
&& \mbox{and} ~ \forall n_1 \in \mathcal{Z}~ \mbox{and} ~n_2 \in
\mathcal{Z} \quad
\mbox{(integers positive or negative)} \nonumber \\
&&\mbox{we have } n_1 \omega_1+n_2 \omega_2 \in  \mathcal{E}
\label{prop1}
\end{eqnarray}
\begin{enumerate}
\item \textit{Then, either the Lebesgue measure $\mu(\mathcal{E})$
of $\mathcal{E}$ is zero or $\mathcal{E}=\mathcal{R} $ is  the
whole real line.} \item \textit{There exists sets $\mathcal{E}$
with property (\ref{prop1}) with zero Lebesgue measure and which
are uncountable }
\end{enumerate}

Since the measure of the real line is infinite,
\textit{measurable} means  locally measurable, i.e.~for any finite
interval  $I(a,\Delta)=[a,a+\Delta]$ of width $\Delta$, the subset
$ I (a,\Delta)\bigcap \mathcal{E}$ is measurable.

There exists many examples of sets having property (\ref{prop1}):
the set of integer multiples of  a given number, the set of
rational numbers. Such sets are countable and  thus have zero
measure.  We shall give also examples of bigger sets which are not
the real line, but uncountable. This theorem proves that they must
have at least zero Lebesgue measure.

\subsection{Proof of proposition (1)}
\textit{Summary of the proof}:  We assume there exists a
measurable set $\mathcal{E}$ with the above property which is not
the real line and such that there exists an interval in which it
has a non zero Lebesgue measure $\mu(I(a,\Delta)\bigcap
\mathcal{E}) \neq 0$ with $ \mu(I(a,\Delta)\bigcap \mathcal{E})
\leq \Delta$. Then
\begin{itemize}
\item we prove it has a uniform measure: $\mu(I(b,\Delta)\bigcap
\mathcal{E})$ only depends on the width of the interval $\Delta$
and is proportional to this width (statement 1)

\item  for $x \notin \mathcal{E}$ , we consider the sequence of
translated sets $\mathcal{E}_n=\mathcal{E}+nx$ for all integers
$n\geq 0$ and prove that there exists a finite integer $p$ such
that $\mathcal{E}_p=\mathcal{E}_0 =\mathcal{E}$. Then,
$\mathcal{E}_{kp+r}=\mathcal{E}_r$ with $k$ and $0\leq r<p$
integers and $\mathcal{E}_r \bigcap \mathcal{E}_{r^{\prime}}
=\emptyset$ when $r\neq r^{\prime}$ and $0\leq r^{\prime}<p$
(statement 2)

\item we consider the sequence of translated sets
$\mathcal{F}_n=\mathcal{E}+x/p^n$ for $n\geq 0$ and prove they are
all disjoints. Then, $ \mu(I(a,\Delta) \bigcap_{n\geq 1} (\bigcup
\mathcal{F}_n)$ should be infinite which is impossible since it is
bounded by $\Delta$. This contradiction proves the absence of
measurable sets with the required properties and thus proves the
theorem (statement 3).

\end{itemize}

Let us assume $\mathcal{E}$ is a measurable set with property
(\ref{prop1}) and with a non vanishing measure.

\subsubsection{statement 1}

We prove first that $\mu(I(a,\Delta) \bigcap \mathcal{E}) = \alpha
\Delta$ where $\alpha$ is some non vanishing constant between $0$
and $1$.

We choose two numbers $\omega_1 \in \mathcal{E}$ and $\omega_2 \in
\mathcal{E}$ such that $\omega_2/\omega_1$ is an irrational
number. This is always possible because if all $\omega_2/\omega_1$
would be rational, then the set $ \mathcal{E}/\omega_0$ with
$\omega_0 \in \mathcal{E}$ would be a subset of the rational
numbers which is countable. Thus, the measure $\mu( \mathcal{E})$
would be zero which contradicts our initial assumption.

Thus, we can find $\omega_1 \in \mathcal{E}$ and $\omega_2 \in
\mathcal{E}$ with $\omega_2/\omega_1$ irrational. Then, the set of
numbers $\omega_{n_1,n_2}= n_1 \omega_1 +n_2\omega_2 \in
\mathcal{E}$ is dense on the real axis (Weyl theorem). Thus for an
arbitrarily chosen  $y$, we can choose a sequence $\omega_i \in
\mathcal{E}$ in that dense set $\{\omega_{n_1,n_2}\}$ such that
$\lim_{i\rightarrow +\infty} \omega_i = y-x$.

Then, the sequence of translated sets $ \mathcal{E}_{i} =
\omega_{i}+ \mathcal{E}=\mathcal{E}$ is invariant.  We have
obviously by translation
 $\mu(I(x,\Delta)\bigcap \mathcal{E})= \mu(I (x+\omega_i,\Delta)\bigcap
 \mathcal{E}_i)
 = \mu(I(x+\omega_i,\Delta)\bigcap \mathcal{E}) $. Then, for $i \rightarrow +\infty$
 $\mu(I(x,\Delta)\bigcap \mathcal{E})=\mu(I(y,\Delta)\bigcap
 \mathcal{E})$.
Consequently,   $\mu(I(a,\Delta)\bigcap \mathcal{E})$ does not
depend on the origin $a$ of the
 interval $I(a,\Delta)$ but only on its width $\Delta$.

 Then, it  comes out by splitting the intervals $\Delta=\Delta_1+\Delta_2$ into  arbitrary pieces that
 $\mu(I(a,\Delta_1+\Delta_2)\bigcap \mathcal{E})=\mu(I(a,\Delta_1)\bigcap \mathcal{E})
 +\mu(I(a,\Delta_2)\bigcap \mathcal{E})$ which readily implies that
 $\mu(I(a,\Delta) \bigcap \mathcal{E}) = \alpha \Delta$ is proportional to $\Delta$ independently of the
 origin $a$ of the interval.
 \textit{Nota}: It is not trivial that the property $f(x+y)=f(x)+f(y)$ for any $x$ and $y$ implies
 $f(x)$ is linear. However, if $f(x)$ is also monotonous
 (for example increasing) which is obviously true for our function
$ \mu(I(a,\Delta_1+\Delta_2)\bigcap \mathcal{E})$ as a function
of  $\Delta$, the proof that $f(x)=\alpha x$ is linear is
straightforward. First, we prove $f(x)=\alpha x$ when $x$ is
rational and since this function is monotone and continuous, we
also have $f(x)= \alpha x$ for $x$ irrational.

\subsubsection{Statement 2}

Let us assume now that $\mathcal{E} \subset \mathcal{R}$ which
implies  $ \exists x\notin \mathcal{E}$. We consider the
translated set  $\mathcal{E}_1=(\mathcal{E}+x)$. We have
 $\mathcal{E}_1 \bigcap \mathcal{E} = \emptyset $.
 Indeed, if we had a nonempty set as intersection ,
 we should find  $\omega_1 \in \mathcal{E}$  and $\omega_2 \in \mathcal{E}$
 such that $\omega_1+x=\omega_2$ therefore, $x\in \mathcal{E}$, in contradiction
 with the initial assumption on $x$.
The measure of the translated set is identical to that of the
original set  since it has uniform measure:  $\mu(I(a,\Delta)
\bigcap \mathcal{E}_1)=
 \mu(I(a-x,\Delta) \bigcap \mathcal{E}) = \alpha \Delta$

Let us now consider the sequence of translated sets
$\mathcal{E}_2=(\mathcal{E}+2x)$, ...,
$\mathcal{E}_n=(\mathcal{E}+nx)$. If $\mathcal{E}_n \bigcap
\mathcal{E}_m \neq \emptyset$, there exists $\omega_1$ and
$\omega_2 \in \mathcal{E}$, such that $\omega_1+nx = \omega_2 +m
x$. Consequently $(n-m)x =\omega_2-\omega_1 \in \mathcal{E}$. Then
$\mathcal{E}_m$ and $\mathcal{E}_n$ are identical since
$\mathcal{E}_m=\mathcal{E}+m x=\mathcal{E}+\omega_1 -
\omega_2+nx=\mathcal{E}+nx=\mathcal{E}_n$. Conversely if  $(n-m) x
\notin \mathcal{E}$, then $\mathcal{E}_n \bigcap \mathcal{E}_m =
\emptyset$.

Thus, if for any integer $p\geq 1$, $px \notin \mathcal{E}$, the
series of sets $\mathcal{E}_n $ are all disjoint. The union
$\mathcal{U}= \bigcup_{n=0,\infty} \mathcal{E}_n$, is a measurable
set and $\mu(I(a,\Delta) \bigcap \mathcal{U})$ is the sum of the
measure of the disjoint components  $\sum_n \mu(I(a,\Delta)
\bigcap\mathcal{E}_n) = \alpha \Delta \times \infty = +\infty$
which is divergent. This is impossible, because this measure must
remain bounded by the measure of the whole interval, $\Delta$ .
Consequently, the assumption $px \notin \mathcal{E}$ for any $p$
is wrong. There are integers such that $px \in \mathcal{E}$. Let
us consider the smallest integer $p_0>1$ such that $p_0x \in
\mathcal{E}$.

Then, $\mathcal{E}_{p_0}= \mathcal{E}$ and if $n$ is written as
the integer division $n=kp_0+r$ with the rest $0\leq r <p_0$, we
have $\mathcal{E}_n =\mathcal{E}_r$  and for $r\neq r^{\prime}$we
have $\mathcal{E}_r  \bigcap \mathcal{E}_{r^{\prime}} =\emptyset$
.

\subsubsection{statement 3}

 For $x \in \mathcal{E}$ and integer $p\geq 1$, we now consider the infinite
 sequence $y_n= x/p^{n}$ for $n=0,1,2,...$.
  We have $y_n \notin \mathcal{E}$ since $y_n \in \mathcal{E}$ would
 implies $p^n y_n =x \in \mathcal{E}$ which contradicts
 the assumption $x \notin \mathcal{E}$. Otherwise, we have  $y_n - y_m \notin \mathcal{E}$ for any $n\neq m$.
It is $y_n - y_m = x (p^{m-n}-1)/p^m$ . If
 $y_n - y_m \in \mathcal{E}$, we would have  $p^m(y_n - y_m) = x (p^{m-n}-1) \in
 \mathcal{E}$. Assuming $m<n$, this is impossible because $p^{m-n}-1= kp+p-1$ with $k=p^{m-n-1}$ and
$0\leq r=p-1<p$. Consequently the sequence   of  translated sets
$\mathcal{F}_n=(\mathcal{E}+y_n)$
 are all disjoint. Therefore, the measure  $\mu(I(a,\Delta) \bigcap
\bigcup_{n\geq1} \mathcal{F}_n)$ is infinite and cannot be smaller
than $\Delta$
 This contradiction proves that $\mathcal{E}$  cannot have nonvanishing measure if it is not the whole
 real axis.

\subsection{Proof of proposition(2)}

Let us consider a sequence of  positive numbers $\omega_n>0$
($n=0,1,...+\infty$)
 such that $\lim_{n\rightarrow +\infty} \omega_{n+1}/\omega_n =0$.
This property implies that $\omega_n$  goes to zero with $n
\rightarrow \infty$ faster than any exponential, that is for any
$\gamma>0$, there exists $K(\gamma)>0$ such that
$$ 0<\omega_n <K(\gamma) e^{-\gamma n}$$
 $\omega_n=1/n!$ is an example of such a sequence.
Obviously, series $\sum_{n=0}^{+\infty} \omega_n < +\infty$ is
absolutely convergent.

Then, let us now consider the set of real numbers $\mathcal{E}$
defined as
\begin{equation}
x\in \mathcal{E}: x=\sum_{n=0}^{+\infty} m_n \omega_n
\label{uncount}
\end{equation}
where $\{m_n\}$ is any bounded sequence of integers positive or
negative that is there exists an integer $N$ such that $\sup_n
|m_n| =N <+\infty$. Then series (\ref{uncount}) is absolutely
convergent since $\sum_{n=0}^{+\infty} |m_n| \omega_n \leq N
\sum_{n=0}^{+\infty} \omega_n <+\infty$.

Then, we have $\mathcal{E}= \bigcup_N \mathcal{E}_N$ where subsets
$\mathcal{E}_N \subset \mathcal{E}$ are defined as
$$ x\in \mathcal{E}_N: x=\sum_{n=0}^{+\infty} m_n \omega_n$$ with $|m_n|\leq N$
for all $n$. We have $\mathcal{E}_N \subset \mathcal{E}_{N+1}$

Thus, assuming $x\in \mathcal{E}$ and $y\in \mathcal{E}$, there
exists $N_x$ and $N_y$ such that $x\in \mathcal{E}_{N_x}$ and
$y\in \mathcal{E}_{N_y}$. Then for any integers $n_x$ and $n_y$,
we have $z=n_x x+ n_y y \in \mathcal{E}_{|n_x| N_x+|n_y| N_y }$
which implies
 $z \in \mathcal{E}$. Consequently,  $\mathcal{E}$ has property (\ref{prop1}).

We now prove that $\mathcal{E}$ has zero Lebesgue measure, i.e.~
$\mu(\mathcal{E})=0$. Since  $\mathcal{E}$ is a countable union of
subsets  $\mathcal{E}_N$, it suffices to prove that
$\mu(\mathcal{E}_N)=0$ for any $N$.

Giving $x=\sum_{n=0}^{\infty} m_n \omega_n \in \mathcal{E}_N$ and
$p$ an arbitrary positive integer,
 the partial sum $x_p=\sum_{n=0}^{p-1} m_n \omega_n$ takes at most $(2N+1)^p$ values.
The  modulus of  the difference $x-x_p$ is bounded as $|x-x_p| =
\sum_{n=p}^{+\infty} |m_n| \omega_n \leq N R_p$ where
$R_p=\sum_{n=p}^{+\infty} \omega_n$. Then, $\mathcal{E}_N$ is
included in the union of $(2N+1)^p$ intervals of width $2NR_p$
which have a total measure bounded as
 $\mu(\mathcal{E}_N) \leq B_p=2N (2N+1)^p  R_p$.

Since for any $\gamma>0$, there exists $K(\gamma)$ such that
$\omega_n <K(\gamma) e^{-\gamma n}$, we have $R_p < K(\gamma)
e^{-\gamma p}/ (1-e^{-\gamma}) =  K^{\prime}(\gamma)  e^{-\gamma
p}$. Choosing $\gamma > \ln (2N+1)$, the upper bound  $B_p<2N
K^{\prime}(\gamma) \left((2N+1) e^{-\gamma}\right)^p $ goes to
zero  for $p\rightarrow +\infty$ and thus can be made infinitely
small. Consequently, the Lebesgue measure $\mu(\mathcal{E}_N)$ of
$\mathcal{E}_N$ must be zero. which implies that $\mathcal{E} =
\bigcup_{N=0}^{+\infty} \mathcal{E}_N$ has also Lebesgue measure
zero.

For proving that $\mathcal{E}$ is uncountable, it suffices to
prove that one of its subset $\mathcal{F}$  is uncountable. This
set is defined as $x\in \mathcal{F}: x=\sum_{n=p}^{+\infty }m_n
\omega_n$ with $m_n=0$ for $n<p$ and  $m_n=\pm 1$ or $0$.  Since
$\omega_{n+1}/\omega_n$ has been assumed to go to zero for
$n\rightarrow +\infty$, integer $p$ can be chosen such that
$\omega_{n+1} < \omega_n/3$ for any $n\geq p$. Then for any $p\leq
m<n$, we have the strict inequalities
$0<\omega_n<\omega_m/3^{n-m}$. This choice will ensure that  all
numbers in $\mathcal{F}$ are surely distinct.

Considering  $x=\sum_{n=p}^{+\infty }m_n \omega_n \in \mathcal{F}$
and $y=\sum_{n=p}^{+\infty }m_n^{\prime}\omega_n \in \mathcal{F}$
where sequences $\{m_n\} \neq \{m_n^{\prime}\}$ are different,  we
define $q$ as
 the smallest $n$ such that $m_n\neq m_n^{\prime}$. Then, we have
$x-y = (m_q-m_q^{\prime})\omega_q+ \sum_{n>q}
(m_n-m_n^{\prime})\omega_n$. Since  $|\sum_{n>q}
m_n\omega_n|<\sum_{n>q} |m_n|\omega_n < \omega_q \sum_{n\geq1}
1/3^n= \omega_q/2$ and similarly $|\sum_{n>q}  m_n^{\prime}
\omega_n|< \omega_q/2$, we have $|x-y
-(m_q-m_q^{\prime})\omega_q|=| \sum_{n>q}
(m_n-m_n^{\prime})\omega_n| <\omega_q$. Since $m_q$ and
$m_q^{\prime}$ are different, we have $|m_q-m_q^{\prime}| \geq 1$
which implies   that $x$ and $y$ cannot be equal.

Since, the set of infinite sequences $m_n$ with $m_n=\pm1$ or $0$
for $n\geq p$ is obviously uncountable (it can be mapped to the
interval $[0,1]$ where numbers are represented in base $3$),
$\mathcal{F}$ is uncountable as well as  $\mathcal{E}$ which
proves the last statement of proposition (2).

\section{Derivation of the integro-differential equation for time-periodic solutions}
\label{AppD}

Searching for periodic solutions with period $T=2 \pi/\Omega$

\begin{equation}
\bm{q}(t + T; \Omega)=\bm{q} (t; \Omega), \quad u_\nu (t + T;
\Omega) = u_\nu (t; \Omega) \label{C.1}
\end{equation}

for all $t$ and all $\nu$, allows a representation as Fourier
series

\begin{equation}
\bm{q}(t; \Omega)= \sum\limits_{n=-\infty}^\infty \bm{q}_n
(\Omega) e^{i \Omega nt} \,\, , \,\, u_\nu (t; \Omega)=
\sum\limits_{n=-\infty}^\infty u_\nu^{(n)} (\Omega) e^{i \Omega
nt} \quad . \label{C.2}
\end{equation}

Substituting this into eqs.\eqref{anharmeq},\eqref{harmeq} and
eliminating $u^{(n)}_\nu (\Omega)$ leads to

\begin{equation}
- (\Omega n)^2 \bm{q}_n (\Omega) + (V' (\bm{q}(t)))_n (\Omega) -
\lambda^2 \sum\limits_\nu \frac{\bm{C}_\nu (\bm{C}_\nu \bm{q}_n
(\Omega))}{\omega^2_\nu -(\Omega n)^2} =0 \quad. \label{C.3}
\end{equation}

The back Fourier transform of eq.\eqref{C.3} yields

\begin{equation}
\ddot{\bm{q}} (t; \Omega) + V' (\bm{q} (t; \Omega)) - \lambda^2
\int\limits_0^T \bm{G} (t - \tau; \Omega) \bm{q} (\tau; \Omega)
d\tau =0 \label{C.4} \end{equation}

with

\begin{equation}
\bm{G} (t; \Omega) = \bm{G} (t + T; \Omega)
=\sum\limits_{n=-\infty}^\infty \bm{G}_n (\Omega) e^{i \Omega n t}
\label{C.5} \end{equation}

and

\begin{equation}
(\bm{G}_n (\Omega))_{\alpha, \beta} \equiv G_{\alpha, \beta}^{(n)}
(\Omega) = \frac{1}{T} \sum\limits_{\nu} \, \frac{C_{\nu, \alpha}
C_{\nu, \beta}}{\omega^2_\nu - (\Omega n)^2} \quad .\label{C.6}
\end{equation}

Note, that $G^{(n)}_{\alpha, \beta} (\Omega)$ is even in $n$.
Accordingly, $\bm{G} (t; \Omega)$ is symmetric in $t$. From
eqs.\eqref{C.5} and \eqref{C.6} we obtain

\begin{eqnarray}
&& G_{\alpha, \beta} (t; \Omega) = G_{\alpha, \beta} (t + T;
\Omega)= \sum\limits_{n- \infty}^\infty  G^{(n)}_{\alpha, \beta}
(\Omega)
e^{i  \Omega n t} \nonumber\\
&&= G_{\alpha, \beta}^{(0)} + 2 \sum\limits_{n=1}^\infty
G_{\alpha, \beta}^{(n)} (\Omega) \cos \Omega n t \quad.
\label{C.7}
\end{eqnarray}

With \cite{Gradshteyn}

\begin{equation}
\sum\limits_{n=1}^{\infty} \, \frac{\cos nx}{n^2 -\alpha^2}
=\frac{1} {2 \alpha^2} - \frac{\pi}{2} \, \frac{\cos \alpha
(\pi-x)} {\alpha \sin \alpha \pi} \quad, 0 \leq x \leq 2 \pi
\label{C.8}
\end{equation}

and eq.\eqref{C.6} we get from eq.\eqref{C.7}

\begin{equation}
G_{\alpha, \beta} (t; \Omega) =\sum\limits_\nu
\frac{C_{\nu,\alpha} C_{\nu, \beta}}{2 \omega_\nu \sin \pi
\frac{\omega_\nu}{\Omega}} \cos \omega _\nu
\Big(|t|-\frac{T}{2}\Big) \quad . \label{C.9}
\end{equation}

which is identical to $G_{\alpha, \beta} (t; \Omega)$ from
eq.\eqref{kernelp}, taking into account $G_{\alpha,\beta} (-t;
\Omega)=G_{\alpha,\beta} (t; \Omega)$.

\end{document}